\documentclass[5p,times]{elsarticle}

\usepackage[numbers]{natbib}
\usepackage[nolist]{acronym}

\usepackage{fontawesome}
\usepackage{float}
\usepackage{placeins}
\usepackage{hyperref}
\usepackage{multirow}
\usepackage{subcaption}
\usepackage{soul}
\usepackage{color,soul}
\usepackage{graphicx}
\usepackage{tikz}
\usepackage{colortbl} 
\usepackage{threeparttable}
\hypersetup{
    colorlinks=false,
    hidelinks
}

\usepackage{amssymb}
\usepackage{amsmath}

\journal{Future Generation Computer Systems}

\begin{document}
\acrodef{3GPP}{3rd Generation Partnership Project}
\acrodef{5G}{5th Generation Mobile Network}
\acrodef{AI}{Artificial Intelligence}
\acrodef{AMF}{Access and Mobility Management Function}
\acrodef{AUSF}{Authentication Server Function}
\acrodef{AUC}{Area Under the ROC Curve}
\acrodef{B5G}{Beyond Fifth Generation}

\acrodef{CPU}{Central Processing Unit}
\acrodef{CHF}{Charging Function}
\acrodef{CSV}{Comma-Separated Values}
\acrodef{DoS}{Denial-of-Service}
\acrodef{DDoS}{Distributed Denial-of-Service}
\acrodef{DL}{Deep Learning}
\acrodef{DNN}{Deep Neural Network}
\acrodef{DRL}{Deep Reinforcement Learning}
\acrodef{DT}{Decision Tree}
\acrodef{ETSI}{European Telecommunications Standards Institute}
\acrodef{FIBRE}{Future Internet Brazilian Environment for Experimentation}
\acrodef{FL}{Federated Learning}
\acrodef{GNN}{Graph Neural Networks}
\acrodef{HTM}{Hierarchical Temporal Memory}

\acrodef{IAM}{Identity And Access Management}
\acrodef{IID}{Informally, Identically Distributed}
\acrodef{IoE}{Internet of Everything}
\acrodef{IoT}{Internet-of-Things}
\acrodef{IP}{Internet Protocol}
\acrodef{KNN}{K-Nearest Neighbors}
\acrodef{LSTM}{Long Short-Term Memory}
\acrodef{MAE}{Mean Absolute Error}
\acrodef{ML}{Machine Learning}
\acrodef{MLaaS}{Machine Learning as a Service}
\acrodef{MOS}{Mean Opinion Score}
\acrodef{MAPE}{Mean Absolute Percentage Error}
\acrodef{MSE}{Mean Squared Error}
\acrodef{mMTC}{Massive Machine Type Communications}
\acrodef{MFA}{Multi-factor Authentication}
\acrodef{MITM}{Man-in-the-Middle}
\acrodef{NGMN}{Next Generation Mobile Networks}
\acrodef{NWDAF}{Network Data Analytics Function}
\acrodef{NS}{Network Slicing}
\acrodef{N3IWF}{Non-3GPP Interworking Function}
\acrodef{N3IUE}{Non-3GPP User Equipment}
\acrodef{NRF}{Network Repository Function}
\acrodef{NSSF}{Network Slice Selection Function}
\acrodef{N3IWUE}{N3 Interface for Untrusted non-3GPP User Equipment}
\acrodef{OSM}{Open Source MANO}
\acrodef{PCA}{Principal Component Analysis}
\acrodef{PCAP}{Packet Capture}
\acrodef{PCAPs}{Packet Captures}
\acrodef{PoC}{Proof of Concept}
\acrodef{PCF}{Policy Control Function}
\acrodef{QoE}{Quality-of-experience}
\acrodef{QoS}{Quality-of-Service}
\acrodef{RAM}{Random Access Memory}
\acrodef{ReLU}{Rectified Linear Unit}
\acrodef{RF}{Random Forest}
\acrodef{RL}{Reinforcement Learning}
\acrodef{RMSE}{Root Mean Square Error}
\acrodef{RNN}{Recurrent Neural Network}
\acrodef{ROC}{Receiver Operating Characteristic}
\acrodef{SDN}{Software-Defined Networking}
\acrodef{SFI2}{Slicing Future Internet Infrastructures}
\acrodef{SLA}{Service-Level Agreement}
\acrodef{SON}{Self-Organizing Network}
\acrodef{SMF}{Session Management Function}
\acrodef{SDO}{Standards Development Organization}
\acrodef{TQFL}{Trust Deep Q-learning Federated Learning}
\acrodef{UDM}{Unified Data Management}
\acrodef{UE}{User Equipment}
\acrodef{UDR}{Unified Data Repository}
\acrodef{UERANSIM}{User Equipment Radio Access Network Simulator}
\acrodef{UPF}{User Plane Function}
\acrodef{U2R}{User to Root Attack}
\acrodef{VoD}{Video on Demand}
\acrodef{VR}{Virtual Reality}
\acrodef{V2X}{Vehicle-to-Everything}

\begin{frontmatter}



\title{An Intelligent Native Network Slicing Security Architecture Empowered by Federated Learning}


\author[1]{Rodrigo Moreira\corref{cor1}}
\ead{rodrigo@ufv.br}                        

\affiliation[1]{organization={Federal University of Viçosa (UFV)},
                city={Rio Paranaíba},
                state={Minas Gerais},
                country={Brazil}}

\cortext[cor1]{Corresponding author}    

\author[2]{{Rodolfo} S. Villaça}

\ead{rodolfo.villaca@ufes.br}

\author[2]{Moisés R. N. {Ribeiro}}

\ead{moises.ribeiro@ufes.br}
\affiliation[2]{organization={Federal University of Espírito Santo (UFES)},
                city={Vitória},
                state={Espírito Santo},
                country={Brazil}}

\author[3]{Joberto S. B. Martins}

\ead{joberto.martins@animaeducacao.com.br}

\affiliation[3]{organization={Salvador University (UNIFACS)},
                city={Salvador},
                state={Bahia},
                country={Brazil}}

\author[4]{João Henrique Corrêa}

\ead{joaocorrea@ufc.br}

\affiliation[4]{organization={Federal University of Ceará (UFC)},
                city={Itapajé},
                state={Ceará},
                country={Brazil}}

\author[5]{Tereza C. Carvalho}

\ead{terezacarvalho@usp.br}

\affiliation[5]{organization={University of São Paulo (USP)},
                city={São Paulo},
                state={São Paulo},
                country={Brazil}}

\author[7]{Flávio de Oliveira Silva}

\ead{flavio@di.uminho.pt}



\affiliation[7]{organization={University of Minho (UMinho)},
                city={Braga},
                country={Portugal}}

\begin{abstract}
Network Slicing (NS) has transformed the landscape of resource sharing in networks, offering flexibility to support services and applications with highly variable requirements in areas such as the next-generation 5G/6G mobile networks (NGMN), vehicular networks, industrial Internet of Things (IoT), and verticals. Although significant research and experimentation have driven the development of network slicing, existing architectures often fall short in intrinsic architectural intelligent security capabilities. This paper proposes an architecture-intelligent security mechanism to improve the NS solutions. We idealized a security-native architecture that deploys intelligent microservices as federated agents based on machine learning, providing intra-slice and architectural operation security for the Slicing Future Internet Infrastructures (SFI2) reference architecture. It is noteworthy that federated-learning approaches match the highly distributed modern microservice-based architectures, thus providing a unifying and scalable design choice for NS platforms addressing both service and security. Using ML-Agents and Security Agents, our approach identified Distributed Denial-of-Service (DDoS) and intrusion attacks within the slice using generic and non-intrusive telemetry records, achieving an average accuracy of approximately $95.60\%$ in the network slicing architecture and $99.99\%$ for the deployed slice -- intra-slice. This result demonstrates the potential for leveraging architectural operational security and introduces a promising new research direction for network slicing architectures.
\end{abstract}


\begin{highlights}
\item Embedding security in slicing architectures through ML-Agent and Security-Agents.
\item Assessing security-aware slicing architecture in nationwide testbeds.
\item Advancing intra-slice security with non-intrusive and generic monitoring metrics.
\item Empowering network slicing architecture with federated learning.
\item Assessing distributed ML-Agents handling attack prediction in real-world testbeds.
\end{highlights}


\end{frontmatter}



\section{Introduction}\label{sec:introduction}

Over the last 40 years, mobile networks have evolved and benefited our society, changing the way we perform daily tasks with applications that are now indispensable, smarter, and more useful than before. This impact is measured when we look at the forecast of having 3.5 billion users consuming connectivity and supporting use cases with more than 70 different industrial segments~\cite{Phyu2023, Habibi2023}. The evolution of mobile networks has been marked by a paradigm shift from conventional networks to software-defined intelligent networks facilitated by softwarization, cloudification, and Network Slicing (NS) paradigms~\cite{Alwis2023, Donatti2023}. These advancements have been made to meet the needs of users and modern applications.

Applications such as \ac{VR}, \ac{IoE}, and Autonomous Vehicles are evolving, requiring, rather than network connectivity, a portion of network resources to cope with specific requirements~\cite{moreira2020, alwis2024}. \ac{NS} involves tailoring a physical network to specific applications and services using three primary baselines: isolation, end-to-end connectivity, and application-driven requirements~\cite{Moreira2021, Fdida2022}. In this context, challenges arise from intelligent and secure network slicing, such as high automation, programmability, interoperability, data orchestration, and zero-touch management~\cite{Bolla2023, Noe2023}. Disruptive paradigms, such as Artificial Intelligence as a Service (AIaaS)~\cite{RodriguesMoreira2023} or \ac{MLaaS}~\cite{Oliynyk2023}, can evolve network slicing architectures that provide security skills, not as a feature to be developed apart from slicing architecture, but rather as native-aware security defenses for network slicing, particularly in architectural core operations and transactions~\cite{Dangi2022, Alwis2023, Abbas2023}. These technologies and findings are essential for realizing network slicing. The evolution of mobile networks is envisioned to integrate land, sky, and sea connectivity~\cite{Pivoto2023} because of the heterogeneity of technologies and domains conducive to security issues~\cite{Guo2022}.

In the literature, we found the use of large-scale testbeds to build and validate network slicing architectures, algorithms, and methods. Additionally, advanced testbeds serve as testing grounds for developing and experimenting with innovative network slicing architectures~\cite{Baldin2019, Silva2019, Nathan2019, Wijethilaka2023, Pontes2023}.  Legacy testbeds~\cite{Silva2019, Both2019} were integrated through a reference architecture, and functionalities were presented in the \ac{SFI2} reference Architecture~\cite{Martins2023, Donatti2023, Moreira2023}. However, incorporating these legacies and cutting-edge testbeds to support network-slicing security experimentation poses challenges. Given the heterogeneous integration scenario for network slicing architecture components, we believe a native security-aware architecture can effectively advance such methods.

Several security issues related to network slicing include impersonation, traffic injection, \ac{DoS} tampering, eavesdropping, reply attacks, and interface monitoring~\cite{Martins2023, DeAlwis2024, Singh2024}. This article introduces network slicing architecture enhancement with a native, distributed, and highly scalable security operation method through a combination of \texttt{Security Agents} and \texttt{ML-Agents}. Existing approaches do not fully explore the distributed \texttt{Security Agent} aligned with the \texttt{ML-Agent} as a countermeasure to improve the operational security of network slice architectures and intra-slice with the ability to enhance the attack-handling capabilities for each control-plane core entity~\cite{Mohammad2023, Singh2024}.

In this context, \ac{FL} is a type of distributed learning where multiple devices or systems train models locally on their own data~\cite{Imteaj2022}, sharing model updates only (such as weights or gradients) with a central server at the \texttt{ML-Agent} in the SFI2 architecture. This server aggregates these contributions without accessing the raw data, allowing for the creation of a global model that preserves data privacy and security across distributed sources. Federated Learning offers several advantages over traditional centralized machine learning, particularly in attack detection within network slicing. In a scenario where each tenant within a network slice has access to their flow data and network attack metrics, federated learning becomes a powerful tool for collaborative model development without compromising tenants' privacy.

We validate the effectiveness of our method by embedding federated learning in \texttt{ML-Agents} to supply \texttt{Security Agents} handling \ac{DDoS} and intrusion attacks on network slicing control-plane entities. Our experimentation and validation \ac{PoC} was based on microservices that provide cognitive services to architectural entities in a highly granular, independent, and customizable manner. \texttt{ML-Agents} and \texttt{Security Agents} are organized in Kubernetes sidecar containers that run within a separate service within a pod. The \texttt{Security Agent} works in the same data plane as the pod services of the \ac{SFI2} Architecture core entities that handle threat identification tasks. By contrast, other entities in the control plane of the architecture usually continue their management functions.

The contributions of this study are as follows: (1) A framework for embedding native security into network slicing architectures through \texttt{ML-Agent} and \texttt{Security Agents}; (2) a functional evaluation of a security-native network slicing architecture capable of handling lifecycle operations and intra-slice security threats; (3) an empirical evaluation of a self-adaptive learning architecture based on federated learning for network slicing architectures; and (4) an evaluation of the Security Agent's capacity to handle on-the-fly attack prediction in network slicing architectures on a nationwide testbeds.

The structure of the remainder of this paper is as follows. The topics that relate to this work are discussed in Session \ref{sec:backgroud}. In Section~\ref{sec:related_work}, we contextualize our work within a broader scope, highlighting the unique contributions of this research. The proposed method is presented in detail in Section~\ref{sec:proposed_method}, followed by a description of the experimental setup and results in Section~\ref{sec:experimental_evaluation}. Section~\ref{sec:concluding_remark} discusses concluding remarks and future directions.

\section{Background}\label{sec:backgroud}

\textbf{Network slicing.} Resource sharing enabled by computing virtualization has influenced mobile networks, particularly in \ac{5G} development mainstream. This has resulted in numerous initiatives from industry, \acp{SDO}, and academia to share network resources, known as Network Slicing (NS)~\cite{Gang2019, Moreira2021}. Network slicing is defined by different standard bodies, such as \ac{NGMN}, as a division of a physical network into multiple networks with capabilities and characteristics oriented to a use case~\cite{ngmn2016}. \ac{3GPP} defines network slicing as a technology that allows the operator to create and customize the network to meet different market demands~\cite{3gpp-ts-28.530}. Based on these pillars, the state of the art has been building and evolving solutions, architectures, and management approaches to offer tailored network resources to users despite the challenges of seamless isolation, performance, and security~\cite{park2023}.

\begin{figure}[H]
  \centering
  \includegraphics[width=0.9\columnwidth]{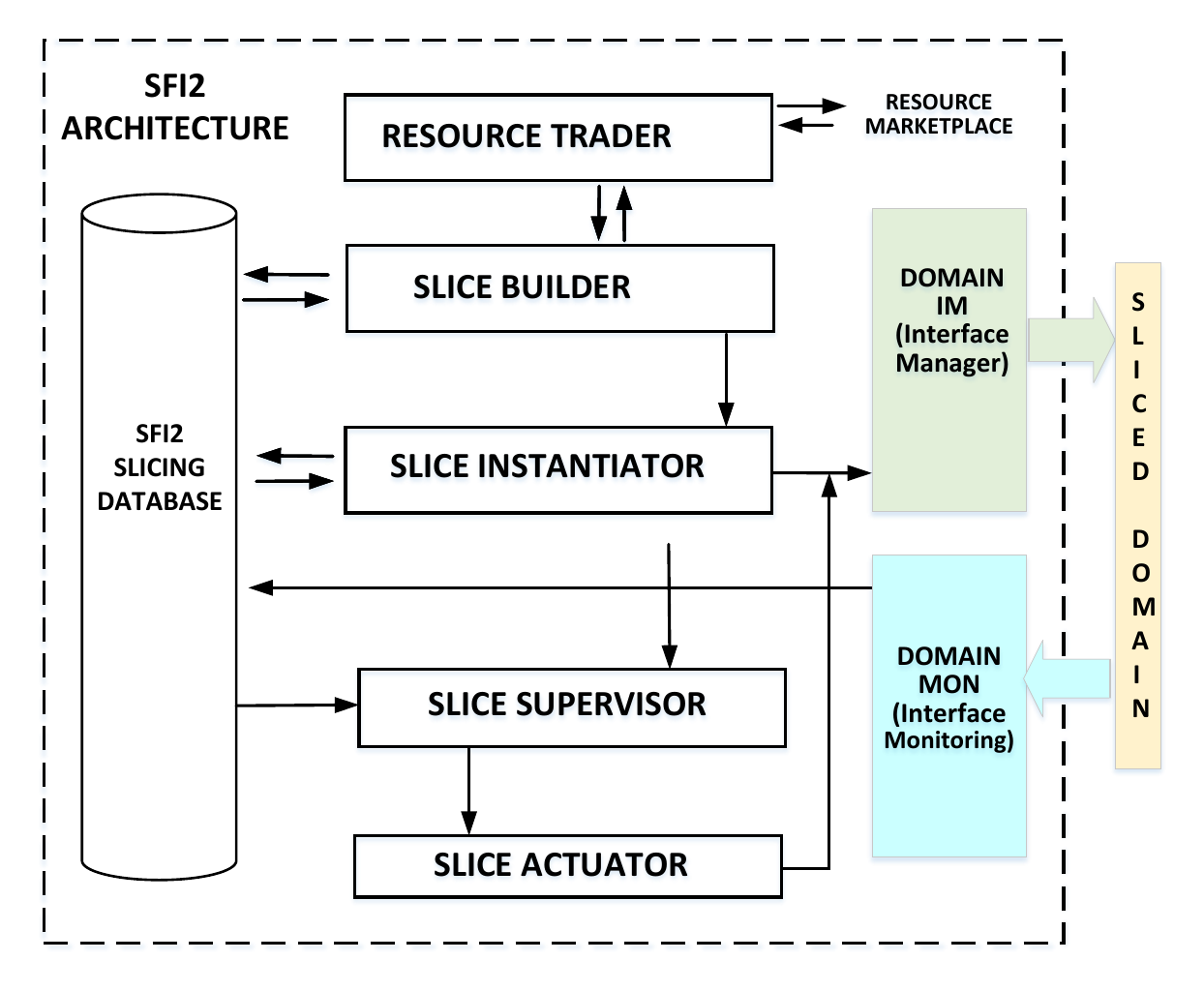}
  \caption{Framework architecture of basic SFI2 building blocks.}
  \label{fig:general_sfi2_architecture}
\end{figure}

\textbf{NS Arquitectures.} A plethora of network slicing architectures for specific domains, industry verticals, and connectivity requirements have emerged~\cite{Parada2018, aloizio2019, Barakabitze2020, Clayman2021, Donatti2023}. While each architecture has its features, they share common aspects, such as the management and lifecycle control loop for network slicing. These architectures provide entities for network slices' preparation, commissioning, operation, and decommissioning phases. The control mechanisms for these phases may vary among architectures; however, coordination, privacy, energy efficiency, and security are still critical, especially for business verticals that utilize network slices \cite{Wijethilaka2022}. Thus, we proposed a conceptual architecture of network slicing \ac{SFI2} focused on security, sustainability, and experimental network integration~\cite{Martins2023}.

\textbf{SFI2 Framework.} The \ac{SFI2} architecture is illustrated in Fig.~\ref{fig:general_sfi2_architecture}, in which the main functional blocks are highlighted~\cite{Martins2023}. The architecture includes a block slice requester, resource trader, slice builder, slice instantiator, slice supervisor, and slice actuator, as well as the interaction with different domains over which network slices can be instantiated. Thus, \ac{SFI2} becomes seamless to deploy network slices in heterogeneous testbeds with different technologies, such as Kubernetes, Docker, or Virtual Machines. The \ac{SFI2} architecture was built to be natively secure and intelligent~\cite{Moreirawpeif2023}. In this paper, we introduce the \texttt{Security Agent}, \texttt{ML-Agent}, with native distributed learning, and \texttt{Monitoring Agent} components~\cite{Martins2023}.

\textbf{Federated Learning}. Federated learning is a machine learning paradigm that trains a robust model by leveraging data spread across heterogeneous devices or servers. Federated learning offers several advantages over centralized approaches for attack detection. Primarily by preserving data privacy and security, tenants can train models locally and share only model updates without exposing sensitive data. It enables collaborative knowledge sharing, allowing tenants to benefit from a robust global model that captures a wider range of attack patterns without directly sharing their data. The approach enhances attack detection by continuously updating the global model with new information from tenants, improving adaptability to emerging threats. Additionally, it reduces bandwidth and computational overhead by minimizing data transfer requirements. It increases resilience to single points of failure by distributing data processing across multiple tenants, where data sensitivity and privacy are critical~\cite{konevcny2016federated, McMahan2017}. The general formula for the federated learning process can be expressed as $F(w)$ = $\frac{1}{K}$ $\times$ $\sum_{k=1}^{K}$ $F_k(w),$ where \(F(w)\) is the global objective function to be minimized, \(w\) represents the model parameters, \(K\) is the total number of clients, and \(F_k(w)\) is the local objective function of the \(k^{th}\) client. Each client computes an update to the model based on its local data, and these updates are then aggregated to form a new global model. The process iterates until convergence or until a satisfactory model is obtained.

\section{Related Work}\label{sec:related_work}

Previous studies have explored various network-slicing approaches, each tailored to specific requirements and applications~\cite{Alwis2023}, including aspects such as security or profit-based resource allocation~\cite{li2024, Gao2024}. Moreover, federated learning has been applied to different network challenges, especially those focused on privacy~\cite{Ma2022, Tusa2023, Babar2024}. Many of these architectures incorporate \ac{AI} capabilities, whereas others employ \ac{AI} for orchestration and other purposes in network slices~\cite{Kaur2022}. This section presents research on applying computational intelligence techniques in network experimental testbeds and using machine learning algorithms to enhance security in network-slicing architectures.

\subsection{Federated Learning for Testbeds}\label{subsec:distributed_learning}

Wijethilaka et al.~\cite{Wijethilaka2022} developed a federated learning strategy to enhance network slicing security, a technology that allocates dedicated logical networks to diverse applications. This manuscript introduces a framework called FLeSO, which utilizes federated learning to train machine learning models to detect anomalies and cyber-attacks in the control plane of sliced networks. The FLeSO framework was tested using an experimental testbed of sliced networks built with open-source tools and an NSL-KDD intrusion dataset. The study indicates that FLeSO is good at identifying attacks, keeping data private, and enabling proactive security measures.

Boualouache et al.~\cite{Boualouache2022} presented a solution for detecting inter-slice attacks in \ac{V2X} 5G networks, which pose a significant threat to the isolation and privacy of network slices. The proposed approach utilizes federated and deep learning to deploy virtual security functions within network slices, which cooperate to train and refine attack detection models. The effectiveness of this method was validated through extensive experimentation conducted on a testbed consisting of sliced networks and the CSE-CIC-IDS2018 intrusion dataset~\cite{Sharafaldin2018}. Our process results in high accuracy in detecting attacks in network slicing management transactions while ensuring data privacy.

Saad et al.~\cite{Saad2023} presents a timely contribution to the field of federated learning in \ac{B5G} networks with zero-touch management, focusing on the pressing issue of poisoning attacks that can prevent the optimal functioning and security of deep learning models utilized in the automated management and orchestration of network slices. To address this challenge, the authors introduce a novel framework called \ac{TQFL}, which employs deep reinforcement learning to select a trusted participant in the federated learning process responsible for detecting and mitigating poisoning attacks using unsupervised learning and dimensionality reduction. The performance of \ac{TQFL} was evaluated through an exhaustive experimentation campaign using the OpenAirInterface platform and a realistic dataset on the latency of the \ac{AMF} function. The results show the effectiveness of \ac{TQFL} in mitigating poisoning attacks while preserving the accuracy and privacy of federated learning models.

\begin{table*}[h]
\centering
\caption{Prior state-of-the-art works towards Security and AI for Network Slicing Architectures.}
\label{tab:sota}
\resizebox{\textwidth}{!}{%
\begin{tabular}{cccccccc}
\hline \hline
\textbf{Approach}                       & \textbf{\begin{tabular}[c]{@{}c@{}}Network \\ Segment\end{tabular}} & \textbf{\begin{tabular}[c]{@{}c@{}}Standards \\ Compatibility\end{tabular}} & \textbf{\begin{tabular}[c]{@{}c@{}}Security \\ Level\end{tabular}} & \textbf{\begin{tabular}[c]{@{}c@{}}Training \\ Paradigm\end{tabular}} & \textbf{Dataset}                                                & \textbf{\begin{tabular}[c]{@{}c@{}}Pro-active \\ Security/Action\end{tabular}} & \textbf{\begin{tabular}[c]{@{}c@{}}Defense/Action \\ Strategy\end{tabular}}                                                                                                                                                                                      \\ \hline \hline
Wijethilaka et al.~\cite{Wichary2022}                            & Slicing Architecture                                                & None                                                                        & Architecture                                                       & Local                                                                 & \multicolumn{1}{c}{None}                                       & \faCircleO                                                                      & \begin{tabular}[c]{@{}c@{}}Manage the resources and services \\ of each slice to guarantee data \\ confidentiality, integrity, and availability.\end{tabular}                                                                                                \\ \hline
Khan et al.~\cite{Khan2022}                             & RAN                                                                 & \ac{3GPP}                                                                        & intra-Slicing                                                         & Local                                                                 & CIC IDS 2017                                                  & \faCircleO                                                                      & \begin{tabular}[c]{@{}c@{}}A centralized controller that \\ coordinates the agents in \\ the slices collects and analyzes the \\ network data, and sends alerts in the \\ event of anomalies.\end{tabular}                                                   \\ \hline
Niboucha et al.~\cite{Niboucha2023}                         & Core                                                                & \ac{3GPP}                                                                        & intra-Slicing                                                         & Local                                                                 & Proprietary                                                     & \faCircleO                                                                      & \begin{tabular}[c]{@{}c@{}}A   deep neural network is \\ used to autonomously detect \\ and mitigate DDoS attacks on \\ the mMTC network slices.\end{tabular}                                                                                                \\ \hline
\multicolumn{1}{c}{Wen et al.~\cite{Wen2022}}         & \multicolumn{1}{c}{Core}                                           & \multicolumn{1}{c}{None}                                                   & \multicolumn{1}{c}{Arquitecture}                                  & \multicolumn{1}{c}{None}                                             & \multicolumn{1}{c}{None}                                       & \multicolumn{1}{c}{\faCircleO}                                                 & \begin{tabular}[c]{@{}c@{}}Integrating   different security solutions \\ into the testbed, such as firewalls, \\ IDS/IPS,   and VPNs, on demand for the \\ network slices that require it.\end{tabular}                                                      \\ \hline
\multicolumn{1}{c}{Silva et al.~\cite{Silva2021}}       & \multicolumn{1}{c}{Core}                                           & \multicolumn{1}{c}{\ac{3GPP}}                                                   & \multicolumn{1}{c}{Arquitecture}                                  & \multicolumn{1}{c}{None}                                             & \multicolumn{1}{c}{None}                                       & \multicolumn{1}{c}{\faCircle}                                                & \begin{tabular}[c]{@{}c@{}}Prevents   DDoS attacks on the 5G \\ control plane through intelligent \\ resource scaling.\end{tabular}                                                                                                                          \\ \hline
\multicolumn{1}{c}{Chilukuri et al.~\cite{Chilukuri2023}}   & \multicolumn{1}{c}{Core}                                           & \multicolumn{1}{c}{\ac{3GPP}}                                                   & \multicolumn{1}{c}{Architecture}                                  & \multicolumn{1}{c}{Local}                                            & \multicolumn{1}{c}{Proprietary}                                & \multicolumn{1}{c}{\faCircleO}                                                 & \begin{tabular}[c]{@{}c@{}}Using   Self-Organizing Network (SON) \\ and Hierarchical Temporal Memory (HTM)-based  \\  learning, the approach detects and \\ isolates malicious users trying to attack   \\ the 5G core using the XDP technique.\end{tabular} \\ \hline
\multicolumn{1}{c}{Wijethilaka et al.~\cite{Wijethilaka2022}} & \multicolumn{1}{c}{Slicing Architecture}                           & \multicolumn{1}{c}{None}                                                   & \multicolumn{1}{c}{Architecture}                                  & \multicolumn{1}{c}{Distributed}                                      & \multicolumn{1}{c}{NSL-KDD}                                    & \multicolumn{1}{c}{\faCircleO}                                                 & \begin{tabular}[c]{@{}c@{}}Federated learning is used to   \\ train machine-learning models to \\ detect anomalies and attacks in the control   \\ plane of sliced networks.\end{tabular}                                                                    \\ \hline
\multicolumn{1}{c}{Boualouache et al.~\cite{Boualouache2022}} & \multicolumn{1}{c}{Slicing Architecture}                           & \multicolumn{1}{c}{None}                                                   & \multicolumn{1}{c}{intra-Slicing}                                    & \multicolumn{1}{c}{Distributed}                                      & \multicolumn{1}{c}{CSE-CIC-IDS2018}                            & \multicolumn{1}{c}{\faCircleO}                                                 & \begin{tabular}[c]{@{}c@{}}Using   federated learning and deep \\ learning to train attack detection models,   \\ virtual security functions are deployed \\ in network slices that collaborate to \\  update these models.\end{tabular}                       \\ \hline
\multicolumn{1}{c}{Saad et al.~\cite{Saad2023}}        & \multicolumn{1}{c}{RAN}                                            & \multicolumn{1}{c}{\ac{3GPP}}                                                   & \multicolumn{1}{c}{Architecture}                                  & \multicolumn{1}{c}{Distributed}                                      & \begin{tabular}[c]{@{}c@{}}Eurecom AMF Resource \\ Consumption   Dataset\end{tabular} & \multicolumn{1}{c}{\faCircle}                                                & \begin{tabular}[c]{@{}c@{}}Deep reinforcement learning was   \\ used to select a trusted participant in \\ federated learning, which is   \\ responsible for proactively \\ detecting and mitigating \\ poisoning attacks in \ac{B5G}   \\ networks.\end{tabular}   \\ \hline
\multicolumn{1}{c}{Jiang et al.~\cite{Jiang2022}}        & \multicolumn{1}{c}{SDN}                                            & \multicolumn{1}{c}{None}                                                   & \multicolumn{1}{c}{Architecture}                                  & \multicolumn{1}{c}{Local}                                      & \begin{tabular}[c]{@{}c@{}}CERNET\end{tabular} & \multicolumn{1}{c}{\faCircleO}                                                & \begin{tabular}[c]{@{}c@{}}
It employs a closed-loop \\ parameter update mechanism \\ and uses DeepAR, a \\ recurrent neural network \\ model, for probabilistic \\ forecasting in slicing admission. \end{tabular}   \\ \hline
\multicolumn{1}{c}{\textbf{Our Proposal}}        & \multicolumn{1}{c}{Slicing Architecture}                                            & \multicolumn{1}{c}{Any}                                                   & \multicolumn{1}{c}{Architecture}                                  & \multicolumn{1}{c}{Local \& Distributed}                                      & \begin{tabular}[c]{@{}c@{}}CIC IDS 2017 \& 5GAD\end{tabular} & \multicolumn{1}{c}{\faCircle}                                                & \begin{tabular}[c]{@{}c@{}}\texttt{ML-Agents} and \texttt{Security-Agents}   \\ deployed as daemonsets in slicing \\ microservice architectures,   \\ providing security for \\ each slicing control-plane \\ entity and intra-slice.\end{tabular}    \\ \hline \hline
\end{tabular}%
}
\end{table*}


\subsection{Sliced Testbeds and Security}\label{subsec:sliced_testbeds_and_security}

Wichary et al.~\cite{Wichary2022} presented a solution to safeguarding and isolating 5G network slices across multiple layers and domains. The proposed approach uses a security attribute model, which associates security controls with the specific requirements of each slice. The study assessed the effectiveness and practicality of various security measures, categorizing them into eight domains and six isolation classes.

Jiang et al.~\cite{Jiang2022} proposed a DeepAR-based probabilistic forecasting model for admission control in network slicing within Software-defined Networks (SDNs). The model is designed for network segmentation and is compatible with 5G and Software-defined Wide Area Network (SD-WAN) standards. They incorporate \ac{AI} and blockchain technologies to enhance network security. The training paradigm leverages DeepAR, a recurrent neural network model, using real-world historical traffic from the China Education and Research Network (CERNET) to proactively manage slice admissions and prevent congestion. In addition, a closed-loop parameter-update mechanism was employed to optimize resource allocation and improve defense strategies.

Khan et al.~\cite{Khan2022} presents a solution to the challenge of detecting \ac{DoS} and \ac{DDoS} attacks on 5G network slices, which can significantly impact the functionality and performance of the associated services. Their proposed method is based on a \ac{RNN}, which utilizes a recently collected dataset from a simulated 5G network slicing testbed. The model's efficacy was validated through accuracy tests, resulting in a remarkable value of $99.99\%$.

Niboucha et al.~\cite{Niboucha2023} addresses the problem of detection and mitigation of \ac{DDoS} attacks on \ac{mMTC} network slices in 5G networks, which can connect many \ac{IoT} devices. The proposed method is a zero-touch security management solution that uses machine learning to predict, identify, and block malicious devices that generate abnormal traffic. The measure used was the detection rate of the model, which was tested on a EURECOM 5G testbed.

Concerning testbeds for security, Wen et al.~\cite{Wen2022} presented VET5G, an end-to-end virtual testbed for experimenting with security in 5G networks. The proposed method is a container-based platform that emulates mobile devices, RAN, and 5G core networks, supporting programmability and isolation. The measures used were the performance and usability of the testbed, which was evaluated in two attack scenarios and a course project.

Wijethilaka et al.~\cite{Wijethilaka2022} focus on ensuring security in sliced networks, a critical component of future telecommunication systems. To address this issue, the authors propose an architecture for a security orchestrator that can provide tailored security services to different network slices and evaluate its feasibility and performance through experimentation using an OpenStack-based testbed, together with \ac{OSM}, PyTorch, and an NSL-KDD intrusion detection dataset.


The paper by Silva et al.~\cite{Silva2021} focuses on the significant issue of \ac{DDoS} attacks on the 5G control plane, which can compromise service availability and security. The authors propose a new approach, REPEL, an intelligent resource-scheduling strategy that utilizes game theory to combat these attacks. The study examines the efficiency and effectiveness of REPEL through a queuing model and an experimental testbed utilizing a virtualized evolved packet-core prototype.

Chilukuri et al.~\cite{Chilukuri2023} present SENTINEL. This framework utilizes the \ac{SON} paradigm and learning based on \ac{HTM} to protect the 5G core control plane from \ac{DDoS} attacks. The framework detects and isolates malicious users attempting to attack the 5G core using a slice aggregator and \ac{MFA}. The efficacy of SENTINEL is evaluated through experimentation on a 5G testbed and utilization of a semisynthetic dataset of anomalies. The results indicate that SENTINEL maintains high levels of service availability for legitimate users while avoiding the expenditure of additional resources.

We summarize the related works in Table~\ref{tab:sota}, where the \emph{Network Segment} column refers to the type of network slicing performed by the slicing architecture. The \emph{Standards Compatibility} column refers to the compatibility of the slicing architecture with standardizing entities, such as \ac{3GPP} and \ac{ETSI}. The \emph{Security Level} column refers to the type of security feature the network-slicing architecture provides, whether for the network slice service  (intra-slice) or the architecture as a whole. The \emph{Training Paradigm} column refers to the method of training AI mechanisms supported by the slicing architecture. The '\emph{Dataset} column summarizes the datasets of the architectures used in the experimental evaluations. The \emph{Pro-active Security/Action} column refers to how the security mechanisms act in the slicing architecture, being reactive and proactive.

\section{Federated Learning Empowering Sliced Testbeds Security}\label{sec:proposed_method}

Network slicing architectures are designed to meet a specific set of connectivity requirements, based mainly on reference models such as \ac{3GPP} and \ac{ETSI}, among others. In previous work, we presented and validated a reference model for network slicing architectures that addressed aspects not fully covered in the state-of-the-art, such as intrinsic security in the functional blocks of architecture and slicing energy efficiency~\cite {Martins2023, Moreira2023, Moreira2023_Aina}. 

In the literature, there are security approaches for network slices from a connectivity or service perspective, with a predominantly \ac{DoS} family of attacks~\cite{Dangi2022, Mohammad2023, DeAlwis2024}. Furthermore, existing security approaches for network slicing focus on a single type of attack owing to the coupled nature of the slicing architecture design. In this paper, we addressed the operational security of slicing architecture.  Our approach allows a slice service provider to handle multiple attacks if the trained ML model is embedded in the \texttt{Security Agents}. We applied intrinsic security to slicing architecture through self-adaptive learning techniques in microservices to provide security for architecture operations regarding slicing life-cycle management.

\begin{figure*}[h]
  \centering
  \includegraphics[width=0.8\textwidth]{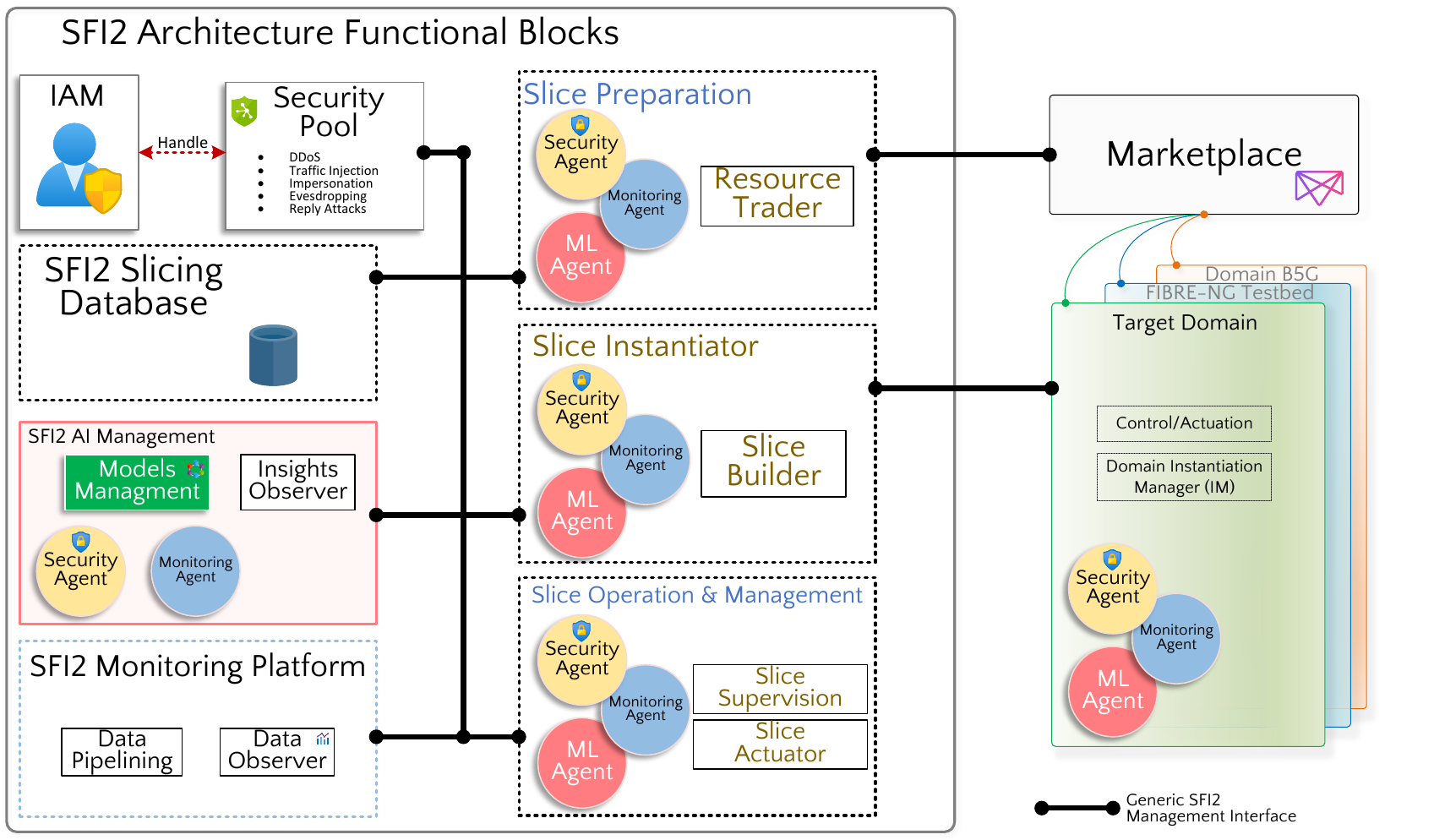}
  \caption{Architecture Building Blocks.}
  \label{fig:architecture}
\end{figure*}

Fig.~\ref{fig:architecture} shows the functional blocks of the SFI2 reference architecture. This architecture envisioned the implementation of network slices considering energy-efficient slicing and integrating experimental networks and testbeds while providing slicing-tailored security functionalities. The functional blocks on the left (\ac{IAM}, Database, \ac{AI} Management, Monitoring, Slice Preparation, Slice Instantiator, Slice Operation \& Management) cooperate at different stages of the life cycle of a network slice to operationalize the connectivity service. The blocks on the right (Control/Actuation and Instantiation Manager) refer to the different target domains where the SFI2 architecture can deploy network slices in the \ac{B5G} domains, \ac{FIBRE} new generation domains~\cite{salmito2014fibre}, well-known experimental testbeds, and others. 

The Marketplace maintains and aggregates different target domains and their resources to enable the deployment of network slices during the slice commissioning phase. In all functional layers of the SFI2 architecture, there is provision for the coexistence of two daemon agents, the Security-Agent and the \texttt{ML-Agent}. These agents act independently and with other functional blocks of the architecture using a microservices approach.

As mentioned, each functional block of the SIF2 architecture has two complementary services: the \texttt{ML-Agent} and the \texttt{Security Agent}, which work asynchronously with the slicing architecture. The \texttt{ML-Agent} performs passive and active functions in the functional block. The passive role involves the prompt and local response to requests for information from AI or analytics associated with the network traffic of the functional block. On the other hand, the active role involves the distributed processing and training of AI models on data common to the functional block it serves, with the \texttt{ML-Agent} periodically reporting the performance of the local model to the SFI2 \ac{AI} Management for aggregation. In line with~\cite{Banerjee2023} findings, to avoid malicious manipulation of system behavior, our approach assumes that all entities in the SFI2 Architecture that support the operation of the network slice life-cycle management have zero trust.

The \texttt{Security Agent} is a critical component of the SFI2 architecture, working as a composite service that operates in parallel with slicing architecture entities. Its main responsibility is actively or passively monitoring the functional block for any security threats, including intrusion and \ac{DDoS} attacks. The \texttt{Security Agent} works in close collaboration with the \texttt{ML-Agent} microservice (which embeds trained AI models into the \texttt{Security Agent}) to detect and prevent malicious traffic patterns in architecture entities, ensuring the overall security and integrity of the SFI2 Architecture. For our implementation, we used the NetData monitoring platform (in {\texttt{Monitoring Agent}}), which can monitor the statistics of both computational nodes and the microservices that run on them; in our case, the Kubernetes DaemonSets.

Our primary aim was to develop a \texttt{Security Agent} that operates as a distributed and asynchronous microservice within our network-slicing architecture. In \ac{5G} networks, \ac{NWDAF} provides subscription and notification services to core entities that consume analytics information from this function. In this work, we extended this indirect approach to providing and consuming services for SFI2 security enforcement. Additionally, we propose a novel method for updating threat defense models using federated learning, enabling entities in the SFI2 Architecture in different phases of the network slice life cycle to handle security actions on the fly.

\begin{figure*}[h]
  \centering
  \includegraphics[width=0.8\textwidth]{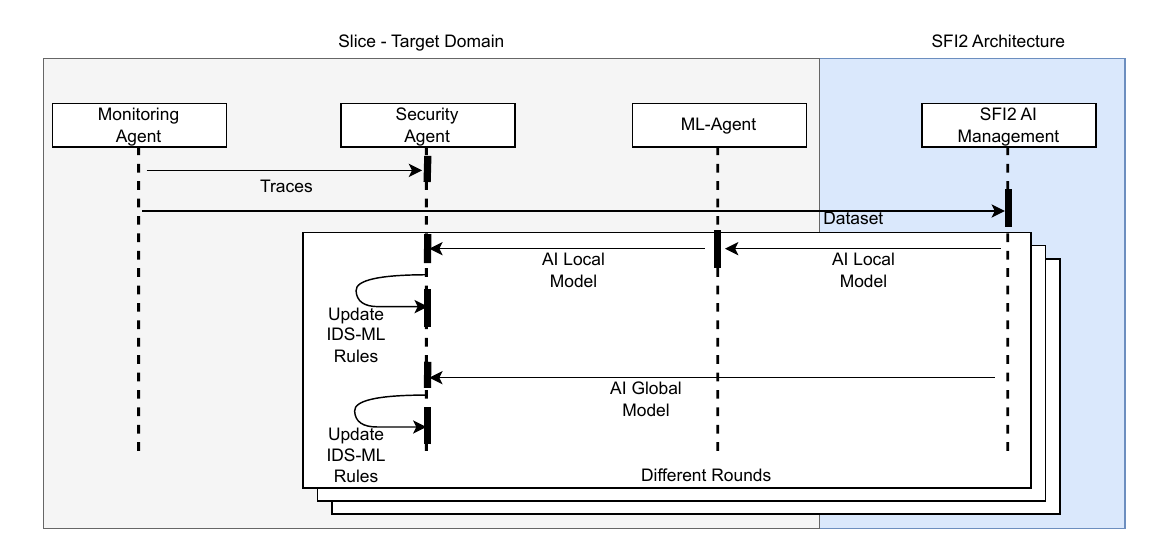}
  \caption{Secure Flow Interactions.}
  \label{fig:flow}
\end{figure*}

An example of this cooperation between the \texttt{Security Agent} and the \texttt{ML-Agent} is constructing an IDS (Intrusion Detection System) based on ML (IDS-ML). Fig.~\ref{fig:flow} illustrates the flow of the interaction between the \texttt{Security Agent}, the \texttt{ML-Agent}, and the \texttt{SFI2 Architecture}. As there will be a \texttt{Security Agent} in each slice created by the SFI2 architecture, it will be responsible for checking any transaction (communication between entities in the slicing architecture) and checking the traffic pattern according to the embedded \ac{AI} model. 

The \texttt{Monitoring Agent} is capable of offering or directing anonymized network flows (traces) such that SFI2 AI Management can store them as a dataset to allow updates to \ac{AI} models in the future. The SFI2 AI Management entity relates to the \texttt{ML-Agents} by inducing these agents to train AI models with global data from SFI2 \ac{AI} Management or with local data, where each \texttt{ML-Agent} is embedded. These traces are transformed into an ML-based input and fed \texttt{ML-Agent}, which generates a local \ac{AI} model related to detecting attacks in the context of that specific slice. The \texttt{Security Agent} updates the IDS-ML rules from this model.

Our slicing architecture extends the literature by allowing \ac{AI} models of each slice (or core entity) to be used to generate a global \ac{ML} model for the entire SFI2 architecture. Our architectural design decision was based on the scalability requirement of the \ac{NS} architecture, so we envisioned a microservice-based architecture with different actuator agents. Each \texttt{\ac{ML}-Agent} sends its \ac{ML} model to \texttt{SFI2 AI Management}, thereby enabling the slicing architecture to handle different \ac{DDoS} threats. The \texttt{SFI2 AI Management} then creates a global \ac{AI} model that considers what has been learned by all \texttt{ML-Agents} scattered by the architecture. Using the global \ac{AI}  model, each slice (or \texttt{Security Agents}) updates the rules in IDS-ML, effectively mitigating potential attacks on the core entity of the slicing architecture. Finally, the \ac{AI} models (local and global) are updated in different rounds and continuously improved over time.

It should be noted that the SFI2 architecture does not have access to the data used to create the \ac{AI} model, only to the \ac{AI} model generated by the \texttt{ML-Agent} coupled with the specific entities of the microservice-based network slicing architecture. Access to this data is limited to the \texttt{Security Agent} and \texttt{ML-Agent} for each slice (or core entity). What is shared with SFI2 AI Management in the SFI2 architecture is only the \ac{AI} model, without containing the data generated by the \ac{AI} model. However, the overall \ac{AI} model comprises contributions from all slices (or core entities) created with information about threats that the core entity itself did not find. In addition, using native federated learning allows the proposal to be placed in the context of Edge Computing with fewer computational resources. This is because the slice (or core entity) can have a more complex \ac{AI} model at the edge without requiring significant computational power to train and generate the model.

\section{Experimental Evaluation}\label{sec:experimental_evaluation}

This paper examines the potential for collaboration between the \texttt{ML-Agent}, \texttt{Security Agent}, and \texttt{Monitoring Agent} to improve the functionality and operational security of network slicing architectures and intra-slice. To achieve this goal, we assessed two (2) experimental perspectives. The first involved evaluating the \texttt{ML-Agent}'s ability to recognize reconnaissance and \ac{DoS} attacks in a sliced 5G core network, using generic non-intrusive monitoring metrics provided by the \texttt{Monitoring Agent} feeding the \texttt{ML-Agent} to identify those threats. The second perspective focuses on detecting anomalies within network slicing architectures' operational components (building blocks), spanning slice preparation, slice implementation, and slice operation and management entities using federated learning.

We have developed a progressive experimental framework, beginning with centralized and classical algorithms to address intra-slice security threats in instantiated and active services. Subsequently, we elevate the analysis of defense mechanisms against security threats by leveraging the unprecedented generalization capabilities of \ac{FL}. This approach enables the construction and validation of slice architectures capable of addressing threats from two perspectives: (1) within the deployed service (intra-slice); and (2) from the operational standpoint of the control plane empowered by \ac{FL}.

\subsection{Intra-Slice Anomaly Detection}\label{subsec:in_service_threats_identification}

In this first experiment, we validate the ability of our architecture to deal with threats involving the running service or deployed network slice. In state-of-the-art, network-slicing architectures, security solutions deal predominantly with the operational security of the architecture, striving to maintain the confidentiality, availability, and integrity of operational components. On the other hand, our proposal sheds light on the security improvement for the service in operation, or intra-slice, by guaranteeing the security aspects for the running network slices on the tenant.

This experiment is based on monitoring a network slice during its operation. We built a \texttt{Monitoring Agent} to collect metrics from the network slice and feed the \texttt{ML-Agent} and the \texttt{Security Agent}. Among the advances in this study, when monitoring the running network slice, we protect it from privacy when the \texttt{Monitoring Agent} inspects the packets' contents, only volumetric and statistical aspects \cite{telemetry2021}. We collected metrics such as network consumption, \ac{CPU}, and memory of running network slice.

\subsubsection{Description of Test}\label{subsubsec:description_of_test}

We validated the feasibility of using generic non-intrusive metrics to assess anomaly detection using basic \ac{ML} algorithms. We employed \ac{KNN}, \ac{DT}, and \ac{RF} to handle the resource consumption dataset to predict anomalies in a running network slice containing a 5G core. Our test aimed to validate the performance of these algorithms for anomaly detection in a 5G core. In contrast, validate the collaboration of \texttt{Security Agent}, \texttt{ML-Agent}, and \texttt{Monitoring Agent}.

\begin{figure}[htbp]
  \centering
  \includegraphics[width=1.0\columnwidth]{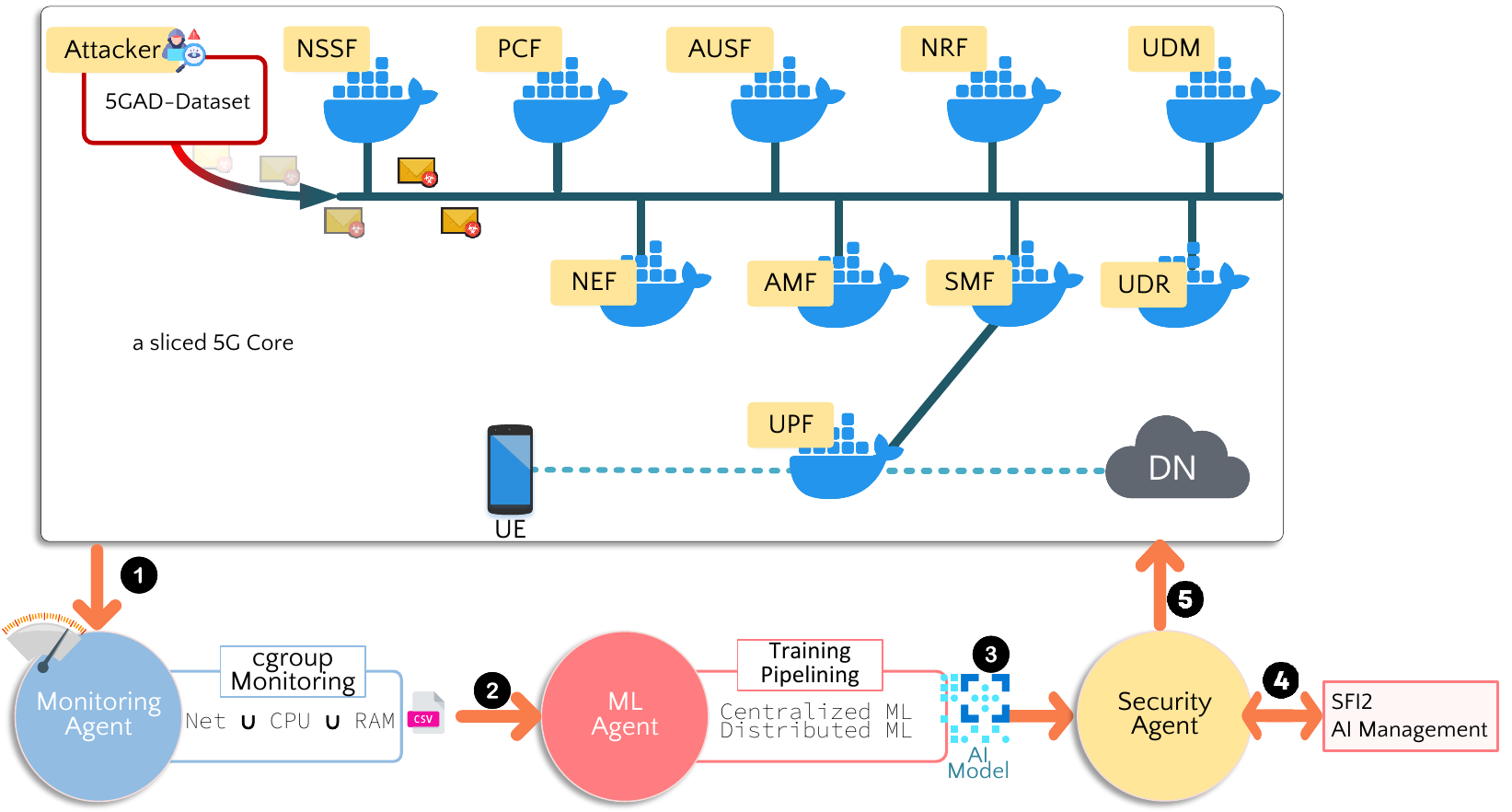}
  \caption{\texttt{Monitoring Agent}, \texttt{ML-Agent} and \texttt{Security Agent} pipelining for intra-slice anomaly detection.}
  \label{fig:in_slice_scenario}
\end{figure}

Fig.~\ref{fig:in_slice_scenario} shows the experimental scenario. Initially, we instantiated an ``Attacker'' container equipped with \textit{\ac{PCAPs}} with traces of attacks on 5G core entities and with the premise of being connected to the 5G core control plane network. Packets are injected into the deployed core using the TCPReplay tool~\cite{tcpreplay}. In phase one (1), the \texttt{Monitoring Agent} is an instance of NetData running in a container that collects different \ac{CPU}, memory, and networking metrics based on the Docker Control Group (cgroup), as detailed in Table~\ref{tab:metrics_and_their_descriptions}. The metric records were transformed into the features of the running slice resource consumption dataset. Some monitored metrics in Table~\ref{tab:metrics_and_their_descriptions}, such as \textit{net\_packets\_eth0}, have additional attributes, including \textit{received}, \textit{sent}, and \textit{multicast}, resulting in a final dataset with 42 features and labels.

\begin{table}[h]
\caption{Generic Slice Metrics and their Descriptions.}
\label{tab:metrics_and_their_descriptions}
\resizebox{\columnwidth}{!}{%
\begin{tabular}{|l|c|}
\hline
\multicolumn{1}{|c|}{\textbf{Metric}} & \textbf{Description}                                                      \\ \hline
cpu                                   & CPU usage by the entity.                                                  \\ \hline
cpu\_limit                            & Maximum allowed CPU usage for the entity.                                 \\ \hline
throttled                             & Number of times the CPU was throttled or restricted.                      \\ \hline
throttled\_duration                   & Total duration of CPU throttling.                                         \\ \hline
mem                                   & Total memory usage by the entity.                                         \\ \hline
writeback                             & Amount of data being written back to the disk.                            \\ \hline
mem\_activity                         & Activity related to memory usage, such as accesses and modifications.     \\ \hline
pgfaults                              & Number of page faults that occurred.                                      \\ \hline
mem\_usage                            & Current amount of memory in use.                                          \\ \hline
mem\_usage\_limit                     & Maximum allowed memory usage.                                             \\ \hline
mem\_utilization                      & Percentage of memory utilization.                                         \\ \hline
mem\_failcnt                          & Count of failed memory allocation attempts.                               \\ \hline
net\_eth0                             & Network traffic on the eth0 interface.                                    \\ \hline
net\_carrier\_eth0                    & Carrier (signal) status of the eth0 network interface.                    \\ \hline
net\_packets\_eth0                    & Number of network packets transmitted and received on the eth0 interface. \\ \hline
net\_errors\_eth0                     & Number of network errors on the eth0 interface.                           \\ \hline
net\_drops\_eth0                      & Network packets dropped on the eth0 interface.                            \\ \hline
net\_fifo\_eth0                       & Number of FIFO errors on the eth0 interface.                              \\ \hline
net\_events\_eth0                     & Network-related events on the eth0 interface.                             \\ \hline
throttle\_io                          & Rate of I/O (input/output) throttling.                                    \\ \hline
throttle\_serviced\_ops               & I/O operations that were throttled.                                       \\ \hline
pids\_current                         & Current number of active processes.                                       \\ \hline
\end{tabular}%
}
\end{table}

Following, as Fig.~\ref{fig:in_slice_scenario} we perform the union operation ($\cup$) based on the timestamp, taking the resulting \ac{CSV} to the \texttt{ML-Agent} in step two (2). The resulting \ac{CSV} in step two (2) is a new resource behavioral dataset employed to validate our method. Different algorithms and \ac{ML} are applied to the data in this phase. The \texttt{ML-Agent} service instance responds to the remote procedure call of \ac{SFI2} AI Management that starts the model training life cycle. We employed classic algorithms such as \ac{KNN}, \ac{RF}, and \ac{DT} to validate our contribution. In phase three (3), when the training of the \ac{AI} models is complete, the Security-Agent receives the trained model. In phase four (4), the trained model weights integrate the SFI2 AI Management model pool, which can serve this model for further requests. In phase five (5), the \texttt{Security Agent} can now perform anomaly detection based on the current running network slicing.

\subsubsection{Dataset}\label{subsubsec:arquitecture_threats_identification}

The 5G Attack Detection (5GAD-2022) dataset consists of intercepted 5G network data, including both normal and malicious traffic in \ac{PCAP} files. The normal data was generated by simulating typical user activities like streaming videos, making web requests, and joining video conferences. The malicious data includes ten types of attacks categorized into reconnaissance, \ac{DoS}, and network reconfiguration. These attacks exploit vulnerabilities in the 5G Core network~\cite{Coldwell2022}.

The dataset was collected in a simulated environment using open-source software free5GC and \ac{UE} simulator~\cite{free5g}. Network traffic was captured using Wireshark, focusing on the application layer to ensure that attack packets were fully included. Each packet was truncated or padded to 1,024 bytes to standardize the data for the machine learning model training~\cite{Coldwell2022}. In our experiment, the \textit{\ac{PCAPs}} were previously processed by changing the source and destination \ac{IP} to enable correct forwarding to core entities.

The dataset consists of \ac{CPU}, memory, and network metrics of all free5GC core entities. Here, \ac{AMF}, \ac{AUSF}, \ac{CHF}, \ac{N3IWF}, \ac{N3IWUE}, \ac{NRF}, \ac{NSSF}, \ac{PCF}, \ac{SMF}, \ac{UDM}, \ac{UDR}, \ac{UE} and \ac{UPF}.

This \textit{\ac{PCAP}} dataset was reinforced in our experimental testbed, leading the \texttt{Monitoring Agent} to record new behavioral resource consumption. We labeled our slice resource consumption dataset according to the original \textit{\ac{PCAP}} dataset. Empirically, we have established a new dataset derived from 5GAD with a proportion of 90\% benign instances and 10\% malignant instances with precision of one second. This was done to simulate a real-world scenario where benign traffic is more prevalent than malignant traffic.

\subsubsection{Evaluation}\label{subsubsec:in_slice_evaluation}

For our evaluation, we used the Fabric testbed, a nationwide testbed on which we deployed a virtual machine with 32GB of memory and 16 cores with an Ubuntu 20.04 operating system, containing scikit-learn, Docker 27.1 and Python 3.11. We started the architecture containers (available here \url{https://github.com/romoreira/SFI2_B5G_Security)}, instantiated the ``Attacker'' node and the sliced 5G core based on free5GC.

We performed 10-\textit{fold} cross-validation on the dataset to ensure that all training data were used as test instances to avoid overfitting. As shown in Table~\ref{tab:security_agent_basic_algoritms_performance}, the classical \ac{ML} models embedded in the \texttt{Security Agent} can identify anomalies in the running network slice (intra-slice). In Table~\ref{tab:security_agent_basic_algoritms_performance}, we compare the performance of the algorithms according to different metrics such as accuracy, F1-Score, recall, and precision.

Accuracy is the proportion of correctly classified instances: $\text{Accuracy} = \frac{TP + TN}{TP + TN + FP + FN}$. Precision measures true positives among predicted positives: $\text{Precision} = \frac{TP}{TP + FP}$. Recall, or Sensitivity, measures true positives among actual positives: $\text{Recall} = \frac{TP}{TP + FN}$. The F1-score, the harmonic mean of precision and recall, is given by $\text{F1-Score} = 2 \times \frac{\text{Precision} \times \text{Recall}}{\text{Precision} + \text{Recall}}$. These metrics offer a comprehensive view of model performance, with F1-score being especially useful for imbalanced datasets. Our results is presented in Table~\ref{tab:security_agent_basic_algoritms_performance}.

\begin{table}[h]
\caption{\texttt{Security Agent} performance in anomaly detection: Values highlighted within the rectangle represent the highest average F1-scores among entities directly impacted by intra-slice attacks.}
\label{tab:security_agent_basic_algoritms_performance}
\resizebox{\columnwidth}{!}{%
\begin{tabular}{l|ccc|ccc|ccc|ccc|}
\cline{2-13}
\textbf{}                               & \multicolumn{3}{c|}{\textbf{Accuracy (\%)}}                                                       & \multicolumn{3}{c|}{\textbf{F1-score (\%)}}                                                       & \multicolumn{3}{c|}{\textbf{Recall (\%)}}                                                         & \multicolumn{3}{c|}{\textbf{Precision (\%)}}                                                      \\ \hline
\multicolumn{1}{|c|}{\textbf{Sliced 5G Core Entity}}         & \multicolumn{1}{c|}{\textbf{DT}} & \multicolumn{1}{c|}{\textbf{KNN}} & \textbf{RF}                & \multicolumn{1}{c|}{\textbf{DT}} & \multicolumn{1}{c|}{\textbf{KNN}} & \textbf{RF}                & \multicolumn{1}{c|}{\textbf{DT}} & \multicolumn{1}{c|}{\textbf{KNN}} & \textbf{RF}                & \multicolumn{1}{c|}{\textbf{DT}} & \multicolumn{1}{c|}{\textbf{KNN}} & \textbf{RF}                \\ \hline
\multicolumn{1}{|l|}{\textbf{AMF}}      & \multicolumn{1}{c|}{99.04}       & \multicolumn{1}{c|}{99.00}        & 99.04                      & \multicolumn{1}{c|}{\tikz[baseline,remember picture]\node[draw=red!100, inner sep=2pt] (rect) {97.14};}       & \multicolumn{1}{c|}{97.00}        & 97.13                      & \multicolumn{1}{c|}{95.27}       & \multicolumn{1}{c|}{94.87}        & 95.08                      & \multicolumn{1}{c|}{99.22}       & \multicolumn{1}{c|}{99.40}        & 99.42                      \\ \hline
\multicolumn{1}{|l|}{\textbf{AUSF}}     & \multicolumn{1}{c|}{92.58}       & \multicolumn{1}{c|}{92.59}        & 92.58                      & \multicolumn{1}{c|}{67.10}       & \multicolumn{1}{c|}{67.02}        & 67.00                      & \multicolumn{1}{c|}{61.83}       & \multicolumn{1}{c|}{61.75}        & 61.74                      & \multicolumn{1}{c|}{94.74}       & \multicolumn{1}{c|}{95.36}        & 95.14                      \\ \hline
\multicolumn{1}{|l|}{\textbf{CHF}}      & \multicolumn{1}{c|}{100}         & \multicolumn{1}{c|}{99.98}        & 100                        & \multicolumn{1}{c|}{100}         & \multicolumn{1}{c|}{99.94}        & 100                        & \multicolumn{1}{c|}{100}         & \multicolumn{1}{c|}{99.90}        & 100                        & \multicolumn{1}{c|}{100}         & \multicolumn{1}{c|}{99.99}        & 100                        \\ \hline
\multicolumn{1}{|l|}{\textbf{N3IWF}}    & \multicolumn{1}{c|}{94.45}       & \multicolumn{1}{c|}{94.49}        & 94.50                      & \multicolumn{1}{c|}{78.47}       & \multicolumn{1}{c|}{78.51}        & 78.62                      & \multicolumn{1}{c|}{71.51}       & \multicolumn{1}{c|}{71.44}        & 91.58                      & \multicolumn{1}{c|}{96.30}       & \multicolumn{1}{c|}{97.01}        & 96.78                      \\ \hline
\multicolumn{1}{|l|}{\textbf{N3IWUE}}   & \multicolumn{1}{c|}{92.75}       & \multicolumn{1}{c|}{92.76}        & 92.78                      & \multicolumn{1}{c|}{68.14}       & \multicolumn{1}{c|}{68.01}        & 68.20                      & \multicolumn{1}{c|}{62.58}       & \multicolumn{1}{c|}{62.45}        & 62.60                      & \multicolumn{1}{c|}{95.49}       & \multicolumn{1}{c|}{96.29}        & 96.10                      \\ \hline
\multicolumn{1}{|l|}{\textbf{NRF}}      & \multicolumn{1}{c|}{100}         & \multicolumn{1}{c|}{99.92}        & 100                        & \multicolumn{1}{c|}{\tikz[baseline,remember picture]\node[draw=red!100, inner sep=2pt] (rect) {100};  }         & \multicolumn{1}{c|}{99.77}        & 100                        & \multicolumn{1}{c|}{100}         & \multicolumn{1}{c|}{99.58}        & 100                        & \multicolumn{1}{c|}{100}         & \multicolumn{1}{c|}{99.96}        & 100                        \\ \hline
\multicolumn{1}{|l|}{\textbf{NSSF}}     & \multicolumn{1}{c|}{92.60}       & \multicolumn{1}{c|}{92.58}        & 92.58                      & \multicolumn{1}{c|}{67.24}       & \multicolumn{1}{c|}{66.74}        & \tikz[baseline,remember picture]\node[draw=red!100, inner sep=2pt] (rect) {66.95};                        & \multicolumn{1}{c|}{61.94}       & \multicolumn{1}{c|}{61.51}        & 61.69                      & \multicolumn{1}{c|}{94.76}       & \multicolumn{1}{c|}{96.21}        & 95.35                      \\ \hline
\multicolumn{1}{|l|}{\textbf{PCF}}      & \multicolumn{1}{c|}{96.44}       & \multicolumn{1}{c|}{96.56}        & 96.53                      & \multicolumn{1}{c|}{87.99}       & \multicolumn{1}{c|}{\tikz[baseline,remember picture]\node[draw=red!100, inner sep=2pt] (rect) {88.27};}        & 88.19                      & \multicolumn{1}{c|}{82.47}       & \multicolumn{1}{c|}{82.35}        & 82.34                      & \multicolumn{1}{c|}{96.65}       & \multicolumn{1}{c|}{97.87}        & 97.63                      \\ \hline
\multicolumn{1}{|l|}{\textbf{SMF}}      & \multicolumn{1}{c|}{100}         & \multicolumn{1}{c|}{99.96}        & 100                        & \multicolumn{1}{c|}{100}         & \multicolumn{1}{c|}{99.88}        & \tikz[baseline,remember picture]\node[draw=red!100, inner sep=2pt] (rect) {100};                          & \multicolumn{1}{c|}{100}         & \multicolumn{1}{c|}{99.79}        & 100                        & \multicolumn{1}{c|}{100}         & \multicolumn{1}{c|}{99.98}        & 100                        \\ \hline
\multicolumn{1}{|l|}{\textbf{UDM}}      & \multicolumn{1}{c|}{92.58}       & \multicolumn{1}{c|}{92.61}        & 92.58                      & \multicolumn{1}{c|}{67.00}       & \multicolumn{1}{c|}{\tikz[baseline,remember picture]\node[draw=red!100, inner sep=2pt] (rect) {67.00};}        & 67.00                      & \multicolumn{1}{c|}{61.74}       & \multicolumn{1}{c|}{61.71}        & 61.74                      & \multicolumn{1}{c|}{95.14}       & \multicolumn{1}{c|}{96.00}        & 95.14                      \\ \hline
\multicolumn{1}{|l|}{\textbf{UDR}}      & \multicolumn{1}{c|}{97.32}       & \multicolumn{1}{c|}{97.29}        & 97.36                      & \multicolumn{1}{c|}{91.26}       & \multicolumn{1}{c|}{91.10}        & \tikz[baseline,remember picture]\node[draw=red!100, inner sep=2pt] (rect) {91.38};                      & \multicolumn{1}{c|}{86.41}       & \multicolumn{1}{c|}{86.07}        & 86.48                      & \multicolumn{1}{c|}{98.10}       & \multicolumn{1}{c|}{98.34}        & 98.32                      \\ \hline
\multicolumn{1}{|l|}{\textbf{UE}} & \multicolumn{1}{c|}{100}         & \multicolumn{1}{c|}{99.99}        & 100                        & \multicolumn{1}{c|}{100}         & \multicolumn{1}{c|}{99.97}        & \tikz[baseline,remember picture]\node[draw=red!100, inner sep=2pt] (rect) {100};                        & \multicolumn{1}{c|}{100}         & \multicolumn{1}{c|}{99.95}        & 100                        & \multicolumn{1}{c|}{100}         & \multicolumn{1}{c|}{99.99}        & 100                        \\ \hline
\multicolumn{1}{|l|}{\textbf{UPF}}      & \multicolumn{1}{l|}{99.95}       & \multicolumn{1}{l|}{99.89}        & \multicolumn{1}{l|}{99.99} & \multicolumn{1}{l|}{99.86}       & \multicolumn{1}{l|}{99.68}        & \multicolumn{1}{l|}{\tikz[baseline,remember picture]\node[draw=red!100, inner sep=2pt] (rect) {99.97};  } & \multicolumn{1}{l|}{99.79}       & \multicolumn{1}{l|}{99.61}        & \multicolumn{1}{l|}{99.95} & \multicolumn{1}{l|}{99.93}       & \multicolumn{1}{l|}{99.75}        & \multicolumn{1}{l|}{99.99} \\ \hline
\end{tabular}%
}
\end{table}

In this experiment, we reinjected packets into a sliced 5G core to trigger reconnaissance, network reconfiguration, and \ac{DoS} attacks. Specifically, we cause \ac{AMF} to generate fraudulent requests to \ac{UDM} while impersonating \ac{AMF}. We also injected the ``Get All Network Functions'' attack without specifying the network function, causing \ac{NRF} to behave in a Byzantine manner. We also triggered a ``Random Data Dump'' that refers to deliberately requesting nf-instances from \ac{NRF}. We triggered ``Automatic Redirect with Timer'' attacks, which caused \ac{UE} traffic to be temporarily redirected by changing policies in \ac{UPF}.

\begin{figure*}[!ht]
    \centering
    
    \begin{minipage}{\textwidth}
        \centering
        \begin{minipage}{0.23\textwidth}
            \includegraphics[width=\textwidth]{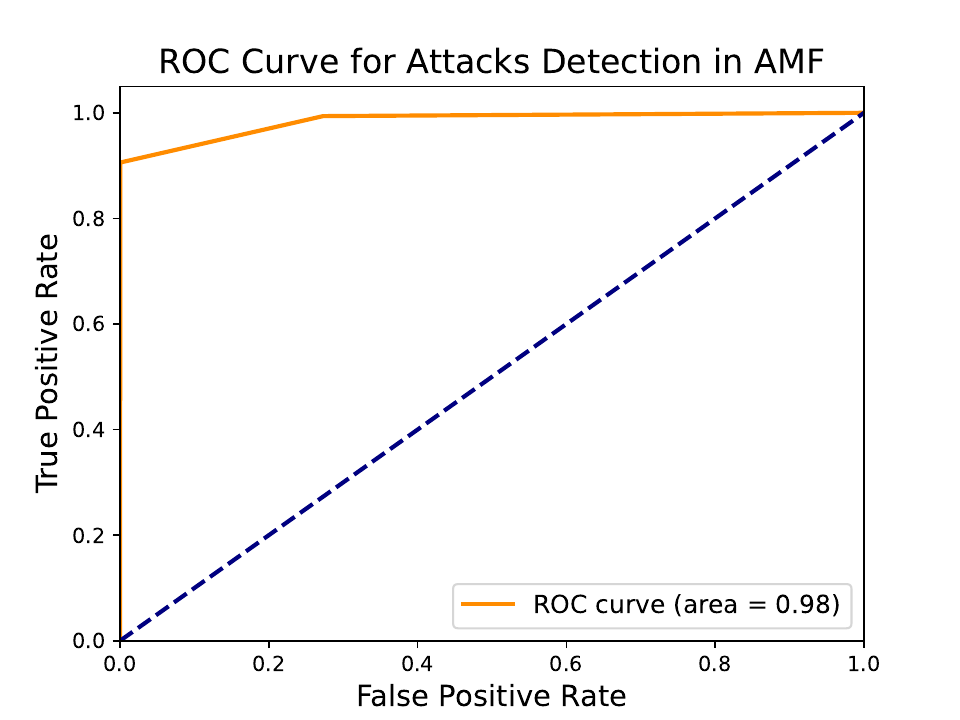}
            \subcaption{}\label{fig:subfig1}
        \end{minipage}
        \begin{minipage}{0.23\textwidth}
            \includegraphics[width=\textwidth]{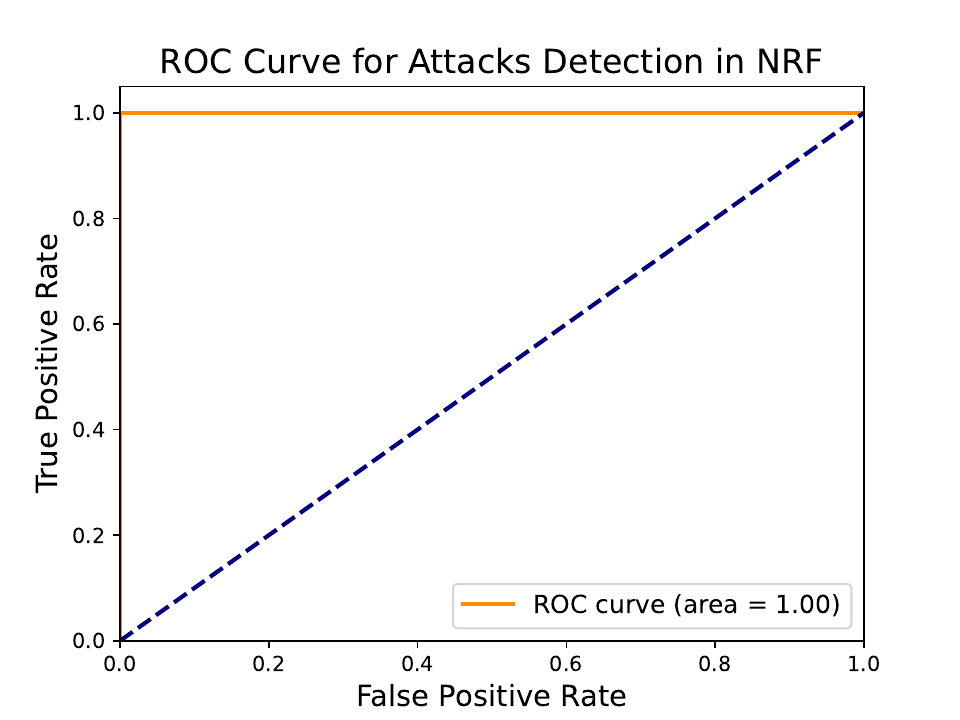}
            \subcaption{}\label{fig:subfig2}
        \end{minipage}
        \begin{minipage}{0.23\textwidth}
            \includegraphics[width=\textwidth]{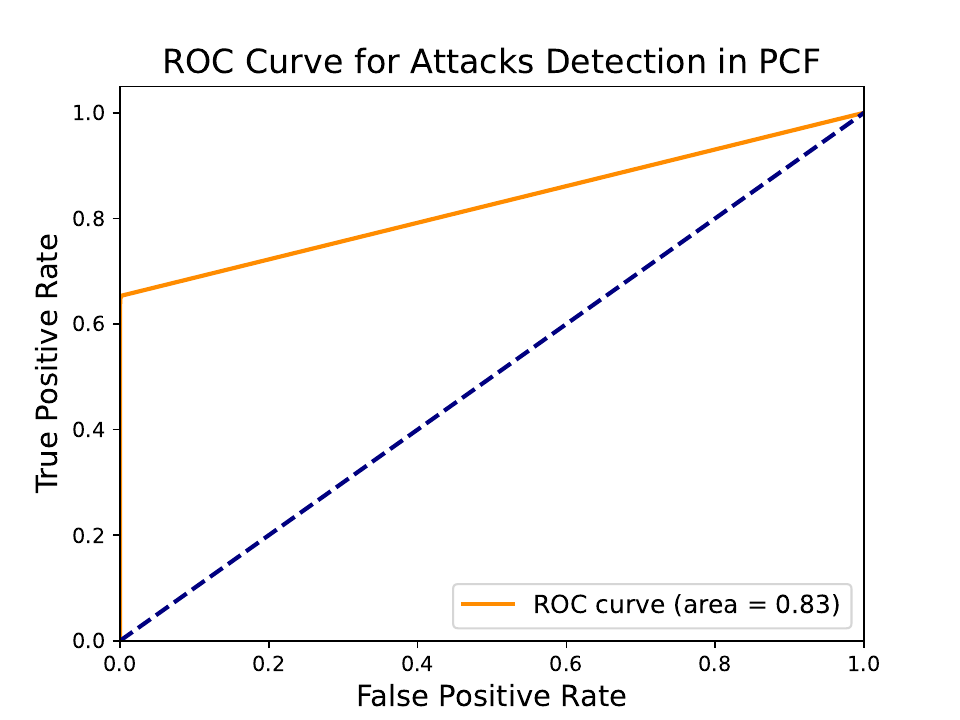}
            \subcaption{}\label{fig:subfig3}
        \end{minipage}
        \begin{minipage}{0.23\textwidth}
            \includegraphics[width=\textwidth]{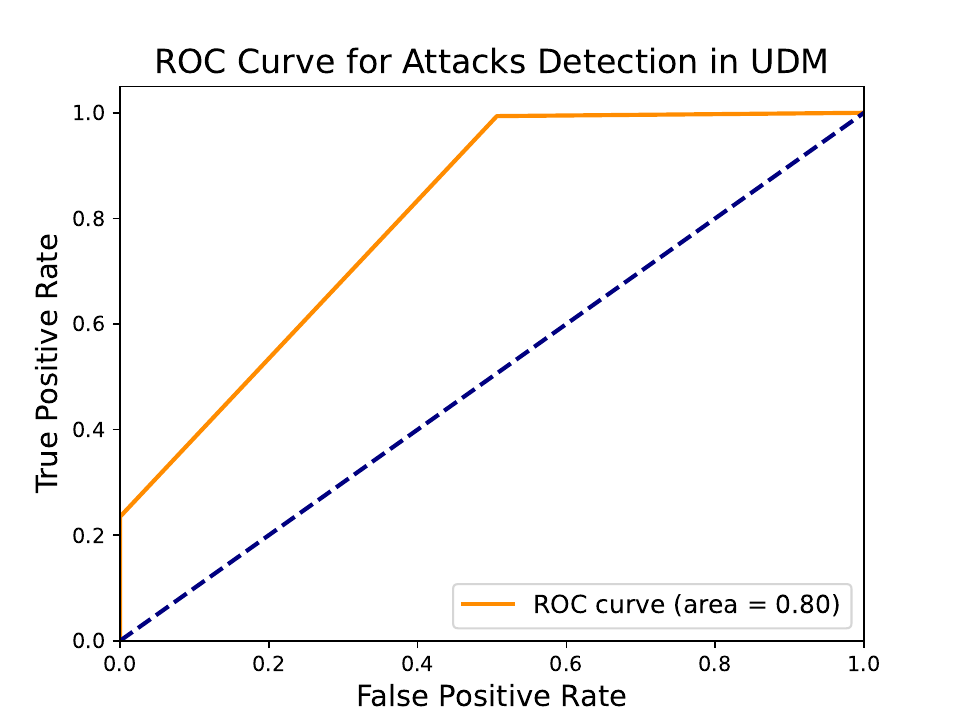}
            \subcaption{}\label{fig:subfig4}
        \end{minipage}
    \end{minipage}
    
    \begin{minipage}{\textwidth}
        \centering
        \begin{minipage}{0.23\textwidth}
            \includegraphics[width=\textwidth]{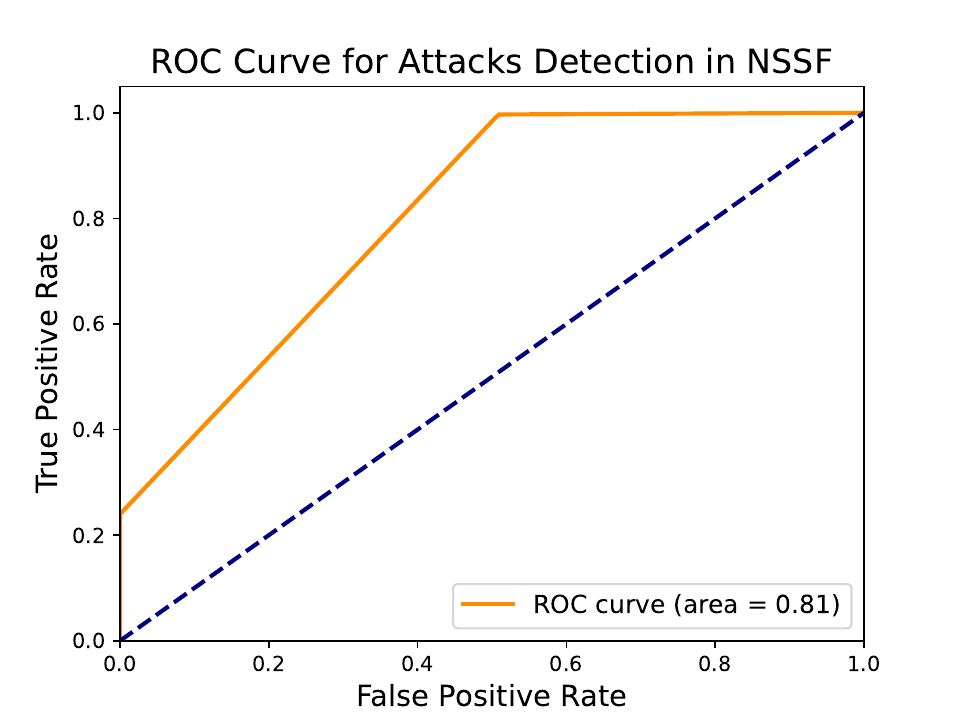}
            \subcaption{}\label{fig:subfig5}
        \end{minipage}
        \begin{minipage}{0.23\textwidth}
            \includegraphics[width=\textwidth]{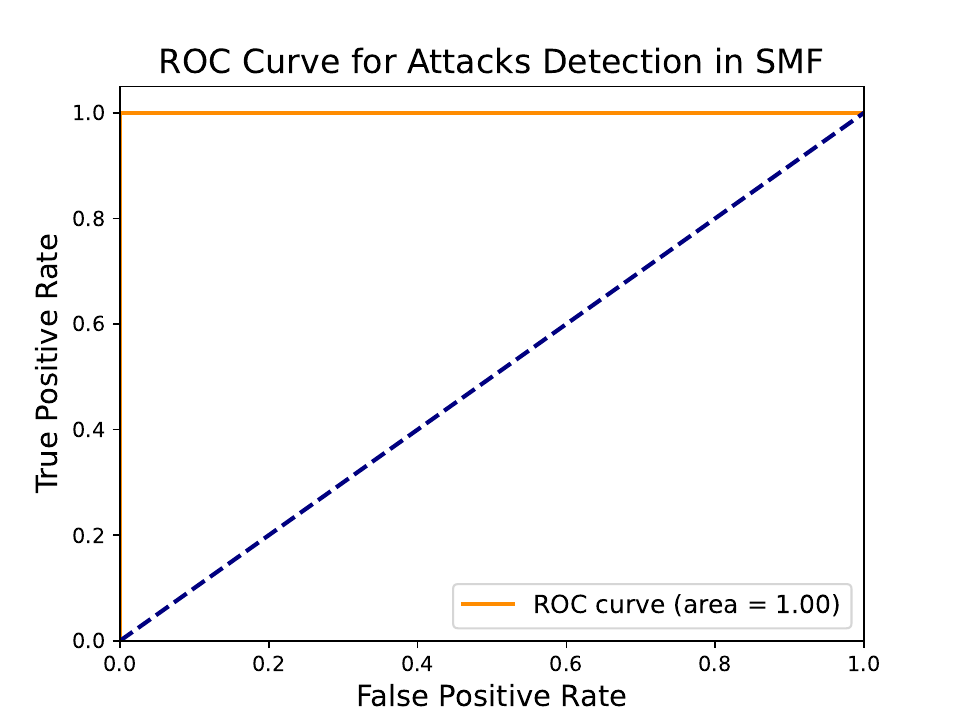}
            \subcaption{}\label{fig:subfig6}
        \end{minipage}
        \begin{minipage}{0.23\textwidth}
            \includegraphics[width=\textwidth]{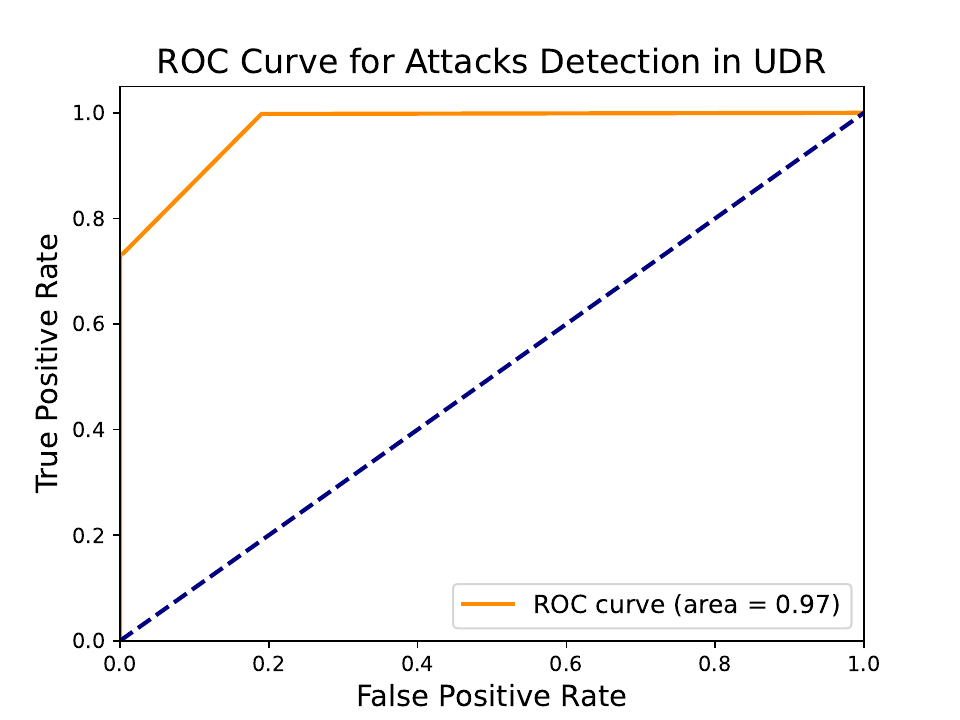}
            \subcaption{}\label{fig:subfig7}
        \end{minipage}
        \begin{minipage}{0.23\textwidth}
            \includegraphics[width=\textwidth]{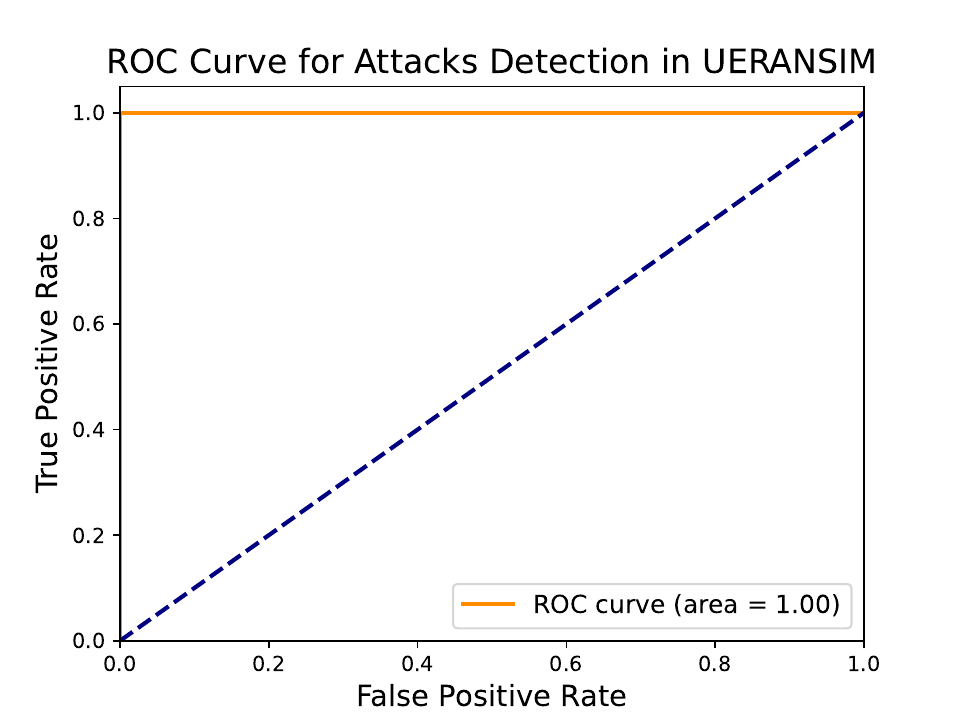}
            \subcaption{}\label{fig:subfig8}
        \end{minipage}
    \end{minipage}
    
    \begin{minipage}{\textwidth}
        \centering
        \begin{minipage}{0.23\textwidth}
            \includegraphics[width=\textwidth]{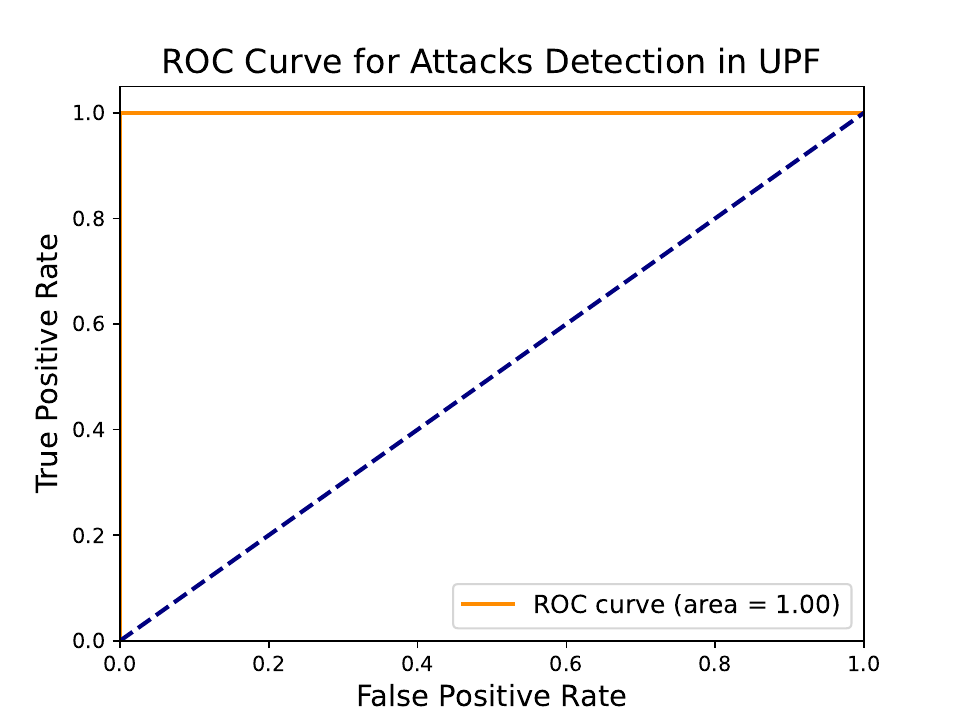}
            \subcaption{}\label{fig:subfig9}
        \end{minipage}
    \end{minipage}
    
    \caption{\ac{ROC} curve for the main entities directly impacted by the intra-slice attacks.}
    \label{fig:roc_curve_for_main_5gc_entities}
\end{figure*}

We inject packets leading to the network reconfiguration phenomenon, thus causing ``Fake \ac{AMF} Delete" causing the sliced 5G core to lose connectivity with the \ac{AMF}. Similarly, we provoke ``Random \ac{AMF} Insert'' scenarios that generate new fraudulent instances of the \ac{AMF}. These events triggered a change in the consumption pattern of the \ac{CPU}, memory, and network resources of the running network slice. Thus, the \texttt{Security Agent} was able to identify these intra-slice nuances, leading to the performance presented in Table~\ref{tab:security_agent_basic_algoritms_performance}.

Finally, we cause a \ac{DoS} attack, specifically the ``Crash \ac{NRF} Attack'' behavior, where malformed requests are deliberately sent to the entity, causing it to fail and triggering abnormal behaviors in the resources of the other entities of the sliced 5G core. In addition to the ``automated drop with timer'' and ``automated redirect with times'' behavior that alternates between redirecting or dropping the traffic of the \ac{UE} in the \ac{UPF}, this leads to anomalous resource consumption in the control entity.

Specifically, the attacks we simulated to evaluate our method directly affected the entities \ac{AMF}, \ac{NRF}, \ac{NSSF}, \ac{PCF}, \ac{SMF}, \ac{UDR}, \ac{UDM}, \ac{UE}, and \ac{UPF}, as listed in Table~\ref{tab:security_agent_basic_algoritms_performance}. Therefore, we observed in more detail the performance of the embedded AI model generated by the \texttt{ML-Agent} combined with the \texttt{Security Agent} on these entities. As shown in Fig.~\ref{fig:roc_curve_for_main_5gc_entities}, the graphs contain the \ac{ROC} curve, showing a trade-off between sensitivity and specificity.

The \ac{AUC} quantifies the overall ability of the model to discriminate between positive and negative classes. An \ac{AUC} value closer to 1 indicates good classification, while a value of 0.5 suggests random guessing. A higher \ac{AUC} value represents a better model performance in distinguishing between classes. The results in Fig.~\ref{fig:roc_curve_for_main_5gc_entities} suggest that the \texttt{Security Agent} combined with the \texttt{ML-Agent} can adequately identify intra-slice anomalies. A higher area under the curve (\ac{AUC}) indicates better performance in distinguishing between legitimate traffic and \ac{DoS} attacks for the \texttt{Security Agent}, reflecting the system's effectiveness in accurately identifying attacks while minimizing false alarms.

\begin{table}[h]
\centering
\caption{Intra-Slice Security Defense Mechanism Comparison.}
\Huge
\label{tab:intra-slice_short_compratison}
\resizebox{\columnwidth}{!}{%
\begin{tabular}{|c|c|c|c|c|c|c|}
\hline
\textbf{Approach}                    & \textbf{Dataset}                         & \textbf{\begin{tabular}[c]{@{}c@{}}Security \\ Threat\end{tabular}} & \textbf{\begin{tabular}[c]{@{}c@{}}Employed \\ Method\end{tabular}}           & \textbf{\begin{tabular}[c]{@{}c@{}}On-time \\ Detection\end{tabular}} & \textbf{\begin{tabular}[c]{@{}c@{}}Low-overhead \\ Monitoring\end{tabular}} & \textbf{\begin{tabular}[c]{@{}c@{}}Defense \\ Efficiency\end{tabular}} \\ \hline
Boualouache   et. al~\cite{Boualouache2022} & CSE-CIC-IDS 2018~\cite{Sharafaldin2018TowardGA} & \ac{DoS}                                                                 & \ac{DL}                                                                            & \faCircleO                                                                    & \faCircleO                                                                          & \begin{tabular}[c]{@{}c@{}}Accuracy: \\ 99.00\%\end{tabular}                                                     \\ \hline
Hossain   et. al~\cite{Hossain2023}         & VeReMi~\cite{heijden2018veremi}                 & \ac{DDoS}                                                                & \begin{tabular}[c]{@{}c@{}}DL with \\ Knowledge \\ Destilation (KD)\end{tabular} & \faCircleO                                                                    & \faCircleO                                                                          & \begin{tabular}[c]{@{}c@{}}Accuracy: \\ 99.00\%\end{tabular}                                                      \\ \hline
Majeed et. al~\cite{Majeed2023}             & CTU-13~\cite{Garcia2014}                        & BotNet                                                              & \ac{DL}                                                                            & \faCircleO                                                                    & \faCircleO                                                                          & \begin{tabular}[c]{@{}c@{}}Accuracy: \\ 97.74\%\end{tabular}                                                      \\ \hline
Boualouache   et. al~\cite{Boualouache2024} & 5G-NIDD~\cite{xtep-hv36-22}                     & \ac{DoS}                                                                 & \ac{FL}                                                                            & \faCircle                                                                   & \faCircleO                                                                          & \begin{tabular}[c]{@{}c@{}}F1-Score: \\ 88.00\%\end{tabular}                                                      \\ \hline
\textbf{Our   Approach}                       & 5GAD~\cite{Coldwell2022}                        & \ac{DoS}                                                                 & Classic \ac{ML}                                                                    & \faCircleO                                                                   & \faCircle                                                                         & \begin{tabular}[c]{@{}c@{}}Accuracy: \\ 100\%\end{tabular}                                                        \\ \hline
\end{tabular}%
}
\end{table}

Table \ref{tab:intra-slice_short_compratison} presents a thorough comparison of existing intra-slice security defense mechanisms, positioning our proposed approach within the current research landscape. The comparison covers key aspects, including the dataset used, type of security threat addressed, employed method, on-time detection capabilities, low-overhead monitoring, and overall defense efficiency. By comparing various approaches, such as those based on \ac{DL}, \ac{FL}, and traditional \ac{ML}, our method achieved 100\% accuracy in detecting \ac{DoS} attacks on the 5GAD dataset through non-intrusive monitoring. This comprehensive analysis underscores the robustness of our approach, particularly in achieving superior detection accuracy without increasing monitoring overhead, thereby contributing to the advancement of security solutions in network slicing.

\subsection{Anomaly Detection in Architecture Building Blocks}\label{subsec:arquitecture_threats_identification}

In this second experiment, we propose a formal evaluation of security-native new advances, showcasing it on the SFI2 Architecture through experiments in which each federated client processes a local dataset~\cite{Sharafaldin2018} consisting of network flows generated by FlowMeter~\cite{FlowMeter2017}. This tool creates tuples of network flows based on the statistical grouping of network packets. Once the packet capture file is generated (\textit{\ac{PCAP}}), it is converted into a \textit{CSV} containing 78 features for each network flow. In this experiment, we validated the ``Security Level'' feature (Table~\ref{tab:sota}) of our architecture based on its ability to handle security threats from an architectural perspective. We have advanced the SFI2 Architecture by empowering each functional entity to manage security threats in the lifecycle control flow on the fly.

We assume that all SFI2 Architecture microservices are ready to handle the network slice lifecycle. In our experimental evaluation, we used \texttt{SFI2 \ac{AI} Management } to trigger a federated learning scenario using \textit{non}–\ac{IID} data. Each federated client (see Fig.~\ref{fig:architecture} as \texttt{ML-Agent\texttt} and associated with microservices as a daemon set) had access to a dataset of a specific type of intrusion and \ac{DDoS} attack. The data in each \texttt{ML-Agent} may not follow the same distribution nor may have the same characteristics as the data on other devices or the overall population, referred to as \textit{non}-\ac{IID} data. This presents challenges for federated learning, such as slow convergence, poor accuracy, and model divergence. 

\begin{figure}[htbp]
  \centering
  \includegraphics[width=0.9\columnwidth]{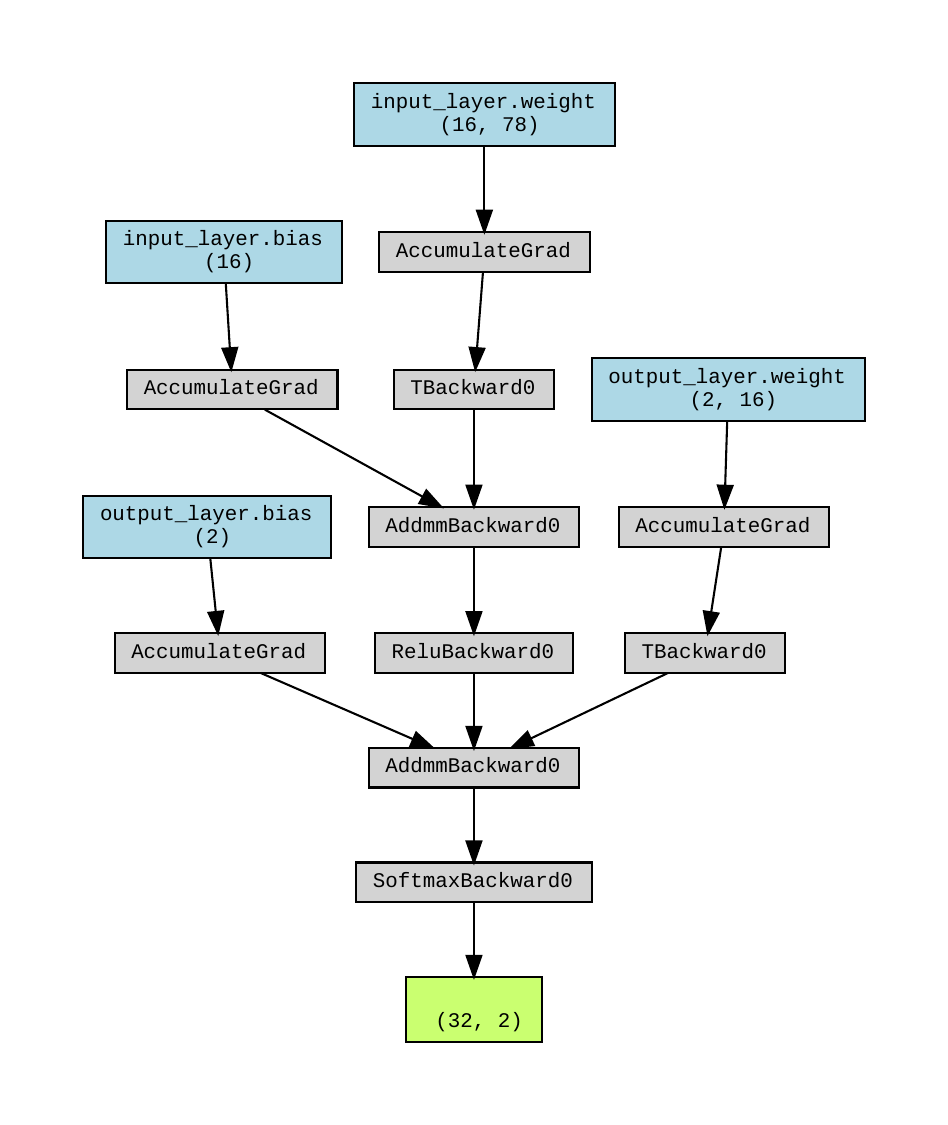}
  \caption{The Neural Network used by each \texttt{ML-Agent}.}
  \label{fig:nn_graph_1}
\end{figure}

We analyzed the optimal hyperparameters for each participant, considering their respective local training sets. In addition, we used the Bayesian approach with the help of the Optuna hyperparameter optimization framework~\cite{Akiba2019}. As depicted in Fig.~\ref{fig:nn_graph_1}, the neural network comprises distinct structures, layers, and input and output data. We sought to minimize loss by considering the number of neural network layers, optimizer, learning rate, and epochs. This neural network architecture was chosen after a previous hyperparameter optimization process, which sought to determine the optimal number of layers to minimize the loss.

We embedded the \ac{LSTM} model shown in Fig~\ref{fig:nn_graph_1} designed for the efficient processing of sequential data in our \texttt{ML-Agent}. The model starts with an input layer that maps 78 input features (detailed in the subsection~\ref{subsubsec:fl_dataset_description}) to 16 units, followed by a hidden layer with the same dimensionality, employing \ac{ReLU} activation, and a dropout layer with a probability of 0.4 to prevent overfitting. The output layer reduces the feature space to two units for binary classification (\ac{DDoS} or \textit{non}-\ac{DDoS}). During training, our model has nodes such as \textit{AccumulateGrad}, \textit{TBackward0}, \textit{AddmmBackward0}, and \textit{SoftmaxBackward0}, which represent operations and gradient accumulations, respectively. For example, \textit{input\_layer.weight} and \textit{output\_layer.weight} with shapes (16, 78) and (2, 16), respectively, are crucial in forward passes, whereas backward operations such as \textit{ReluBackward0} and \textit{SoftmaxBackward0} ensure accurate gradient computation during the backward pass, culminating in the final output of the shape (32, 2).

\subsubsection{Description of Test}\label{subsubsec:description_of_test_scenarios}

We have formalized the evaluation of our SFI2 reference architecture security extension by conducting experiments on a testbed that replicated the production network conditions. We deployed a virtual machine with 32 GB of memory, an 8-core \ac{CPU}, and a GPU RTX 4060 Ti with 8 GB of memory, Ubuntu 20.04 operating system. Flower federated learning framework, cuDNN 12.0 toolkit combined with Torch 2.3. Our dataset consisted of benign and malicious network traffic~\cite{Sharafaldin2018}. The testing process was divided into two phases. The first phase involves offline training of deep neural network algorithms in a local and federated manner. In the second phase, we implemented the learned models in the SFI2 architecture, specifically in the distributed scenario of the testbed, where each \texttt{ML-Agent} runs as microservices in different architectural blocks.

The second experiment involved running the functional blocks of the SFI2 Architecture on different testbed nodes. The SFI2 prediction API can receive a network flow in tuple format and judge its traffic class, benign or malignant. Therefore, we measured the API response time capacity to assess the API readiness regarding the response of our architecture when running production slices.

Finally, validation was performed on a nationwide physical testbed in the Future Internet Brazilian Environment for Experimentation New Generation testbed, which is a microservice-based testbed with many compute nodes spread across educational institutions in Brazil and is designed to be an evolution of the previous FIBRE testbed supported by the National Education and Research Network (RNP)~\cite{salmito2014fibre}. This network is geographically distributed and has an interconnection between Kubernetes nodes via an Internet Protocol (IP) network that connects different research institutions in Brazil.

In our current implementation, we considered only a centralized coordinator within the testbed. This decision was made because the centralized coordinator is located in the AI agent of the SFI2 architecture, which is protected by a security agent. It is designed specifically for and is accessible only to SFI2 tenants, ensuring a secure and controlled environment for federated learning processes. However, we recognize the potential benefits of using blockchains and distributed coordinators for federated learning, particularly in enhancing participant security, transparency, and trust. As such, exploring these approaches represents a valuable direction for future work, where decentralized coordination mechanisms could further strengthen the system's resilience and scalability further~\cite{Juncen2023}.

\subsubsection{Dataset}\label{subsubsec:fl_dataset_description}

We chose four days to train and validate the deep neural networks, encompassing tuples of network flows from different days and times. We used 90\% of the time for training and 10\% for testing, and each experiment was performed 10 (ten) times. Each capture or dataset acquisition day was assigned to a single federated client during the learning process, as listed in Table~\ref{tab:table_label}. The dataset was divided as follows: Monday featured only regular activity (normal traffic with different applications), whereas Tuesday through Friday included hybrid attacks and regular activity. The neural network structure and the layers employed for each federated client are shown in Fig.~\ref{fig:nn_graph_1}. 

\begin{table}[htbp]
\centering
\caption{Dataset file description and distribution.}
\label{tab:table_label}
\scriptsize
\begin{tabular}{cccc}
\hline \hline
\textbf{Day}              & \textbf{\begin{tabular}[c]{@{}c@{}}Size \\      (\# lines)\end{tabular}} & \textbf{\begin{tabular}[c]{@{}c@{}}\% \\      of Malignant\end{tabular}} & \textbf{\begin{tabular}[c]{@{}c@{}}Assigned to \\      which \texttt{ML-Agent}\end{tabular}} \\ \hline \hline
Tuesday                   & 445,909                                                                   & 3.1                   & \#1                                                                                 \\ \hline
Wednesday                 & 692,703                                                                   & 36.48                 & \#2                                                                                 \\ \hline
\multirow{2}{*}{Thursday} & 288,602                                                                   & 0.01                  & \#3                                                                                 \\
                          & 170,366                                                                   & 1.28                  & \#4                                                                                 \\ \hline
\multirow{3}{*}{Friday}   & 286,467                                                                   & 55.48                 & \#5                                                                                 \\
                          & 191,033                                                                   & 1.03                  & \#6                                                                                 \\
                          & 225,745                                                                   & 56.71                 & \#7                                                                                 \\ \hline \hline
\end{tabular}%
\end{table}

\begin{figure*}[h]
    \centering
    \subfloat[]{%
        \includegraphics[width=0.243\textwidth]{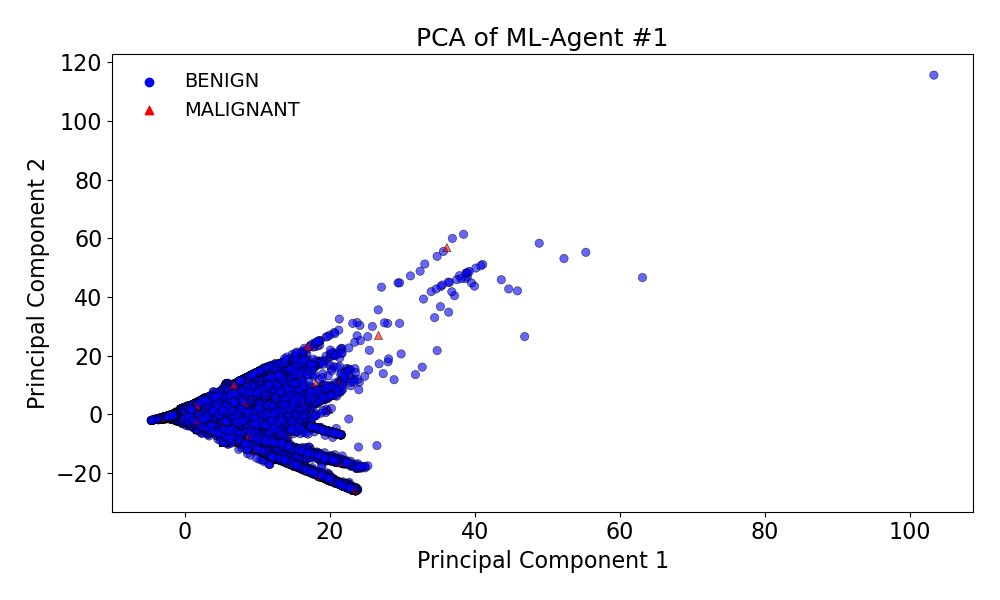}
    }%
    \hfill
    \subfloat[]{%
        \includegraphics[width=0.243\textwidth]{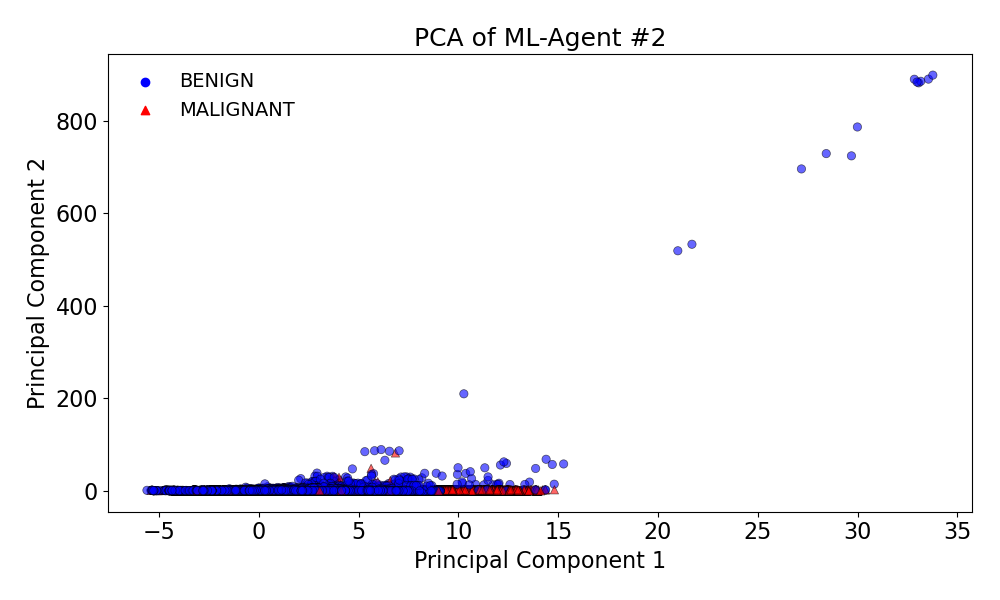}
    }%
    \hfill
    \subfloat[]{%
        \includegraphics[width=0.243\textwidth]{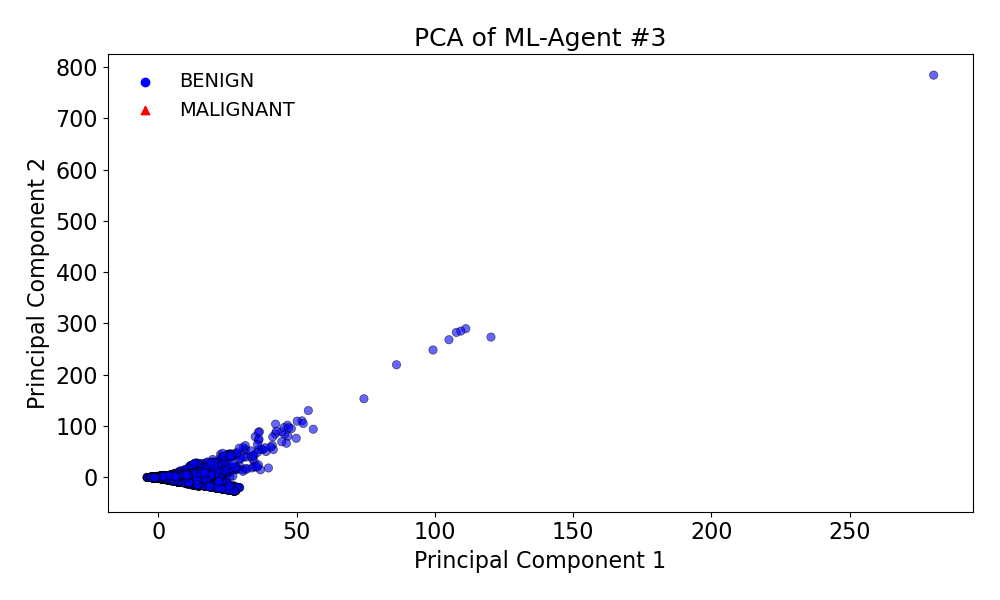}
    }%
    \hfill
    \subfloat[]{%
        \includegraphics[width=0.243\textwidth]{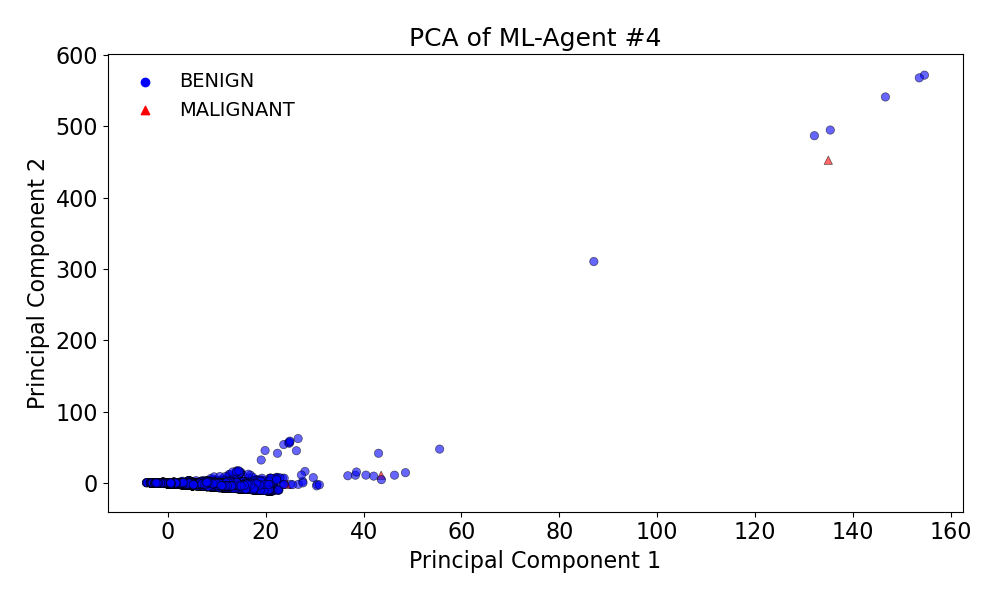}
    }\\
    \subfloat[]{%
        \includegraphics[width=0.28\textwidth]{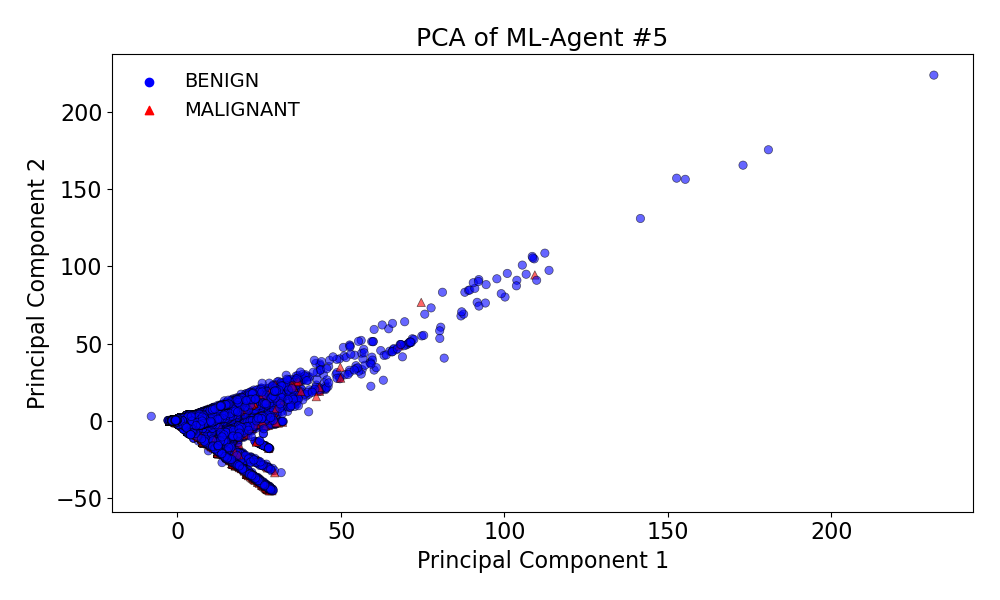}
    }%
    \hfill
    \subfloat[]{%
        \includegraphics[width=0.28\textwidth]{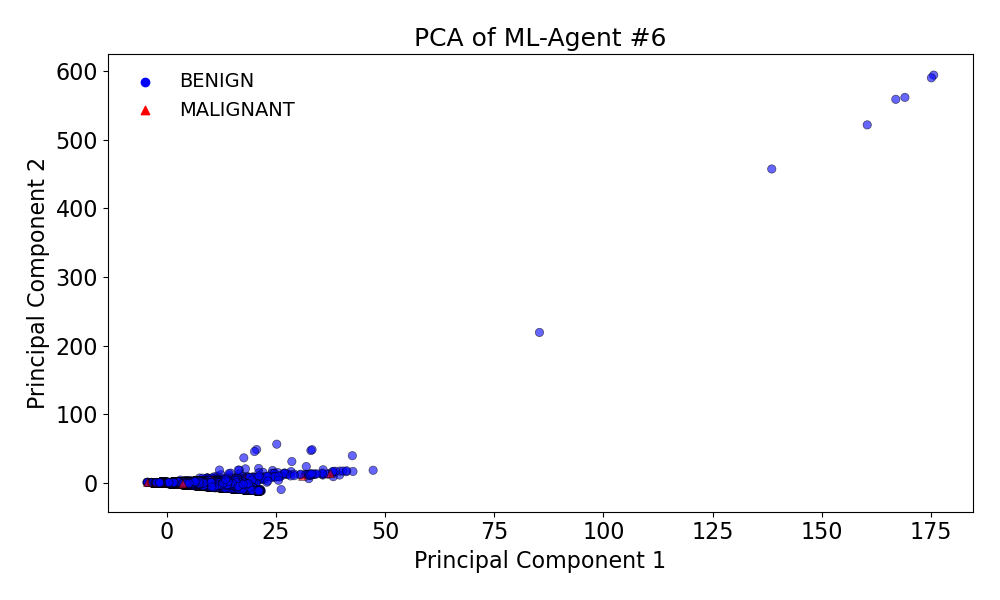}
    }%
    \hfill
    \subfloat[]{%
        \includegraphics[width=0.28\textwidth]{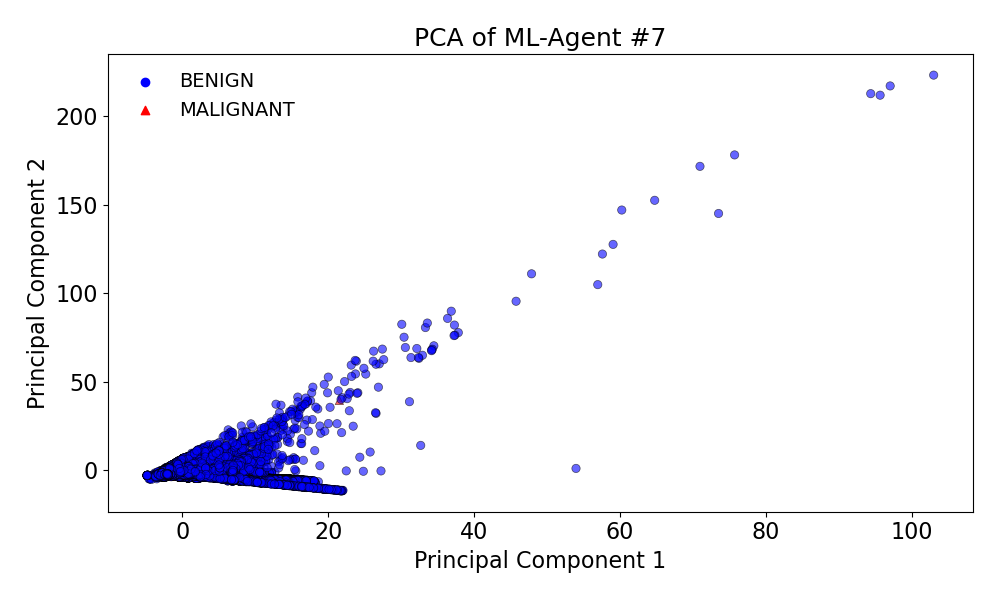}
    }
    \caption{\ac{PCA} of each \texttt{\ac{ML}-Agent} dataset.}
    \label{fig:pca_figures}
\end{figure*}

Within just four (4) days of the existing dataset~\cite{Sharafaldin2018}, time divisions into morning, afternoon, and evening led to the separation of more than four (4) datasets for each \texttt{ML-Agent}, as shown in Table~\ref{tab:table_label} (column ``Assigned to which \texttt{ML-Agent}''). The dataset comprises a real network encompassing various devices such as firewalls, switches, and routers. We previously trained models for the SFI2 AI Management building block, allowing future microservices of the SFI2 reference architecture to import the model. Later, using the trained and operational model, we validated the performance of the security API in classifying network flows from both network slices and functional blocks of the architecture.

The chosen dataset represents the diversity of devices and user behaviors it brings. The dataset was constructed to include various intrusions and benign traffic at different times of the day to capture the unique characteristics of each time slot. The first day of the week, Monday, was excluded from the local training process of our experiment because it consisted of only regular traffic. The remaining days were considered for training, as they contained a mixture of benign and malignant traffic.

To understand the type of data \texttt{\ac{ML}-Agents} have dealt with, we conducted a Principal Component Analysis (PCA), which is a statistical technique used for dimensionality reduction. \ac{PCA} reduces dimensionality by transforming the original variables into a new set of variables, the main components. 
The \ac{PCA}-generated scatter plots reveal that classes have considerable overlap within features, which may lead to difficulties in achieving high accuracy or convergence in terms of learning in certain \ac{AI} models. Fig.~\ref{fig:pca_figures} shows that the malignant and benign classes are mixed, indicating an intrinsic classification challenge for \texttt{\ac{ML}-Agents}. To migrate some of this overlap, we conducted a hyperparameter optimization.

\subsubsection{Analysis Method}\label{subsubsec:analysis_method}

We present the results of optimizing the hyperparameters of each \texttt{ML-Agent} using its local dataset, where the hyperparameters were refined using the Tree-Structured Parzen Estimator (TPE) algorithm~\cite{TPE2011} to maximize the accuracy of each model coupled to the local \texttt{ML-Agent}. Table~\ref{tab:hyperparameters_values} lists the search space and hyperparameters. It is worth noting that the \textit{non}-\ac{IID} format of the dataset managed by each \texttt{ML-Agent} resulted in the discovery of diverse hyperparameters, even when using the same neural network for every \texttt{ML-Agent}. In our experiments, it was necessary to optimize the hyperparameters, as we found that the hegemonic hyperparameters for each \texttt{ML-Agent} and its dataset in the \textit{non}-\ac{IID} scenario prevented the global model from converging its learning rate.

\begin{table*}[h]
\renewcommand{\arraystretch}{1.6}
\centering
\caption{Dataset description and Hyperparameters for each \texttt{ML-Agent}.}
\label{tab:hyperparameters_values}
\scriptsize
\begin{tabular}{ccp{6cm}ccc}
\hline 
\textbf{ML-Agent} & \textbf{Dataset} & \textbf{Attacks}                                                                                                                                                          & \textbf{\begin{tabular}[c]{@{}c@{}}Learning \\ Rate (LR)\end{tabular}} & \textbf{Optimizer} & \textbf{Epochs} \\ \hline 
\textbf{\#1}               & Tuesday          & Brute Force, FTP-Patador, and SSH Patator                                                                                                                                 & 0.0003074258400864182                                                  & Adam               & 10              \\ 
\textbf{\#2}               & Wednesday        & DoS/DDoS: Slowloris, Slowhttptest, Hulk, and GoldenEye & 0.0005025961155459187                                                  & RMSprop            & 10              \\ 
\textbf{\#3}               & Thursday         & Infiltration: Dropbox, Meta exploit Win Vista, Cool disk – MAC, Dropbox download, and Win Vista & 0.00010603472201401003                                                 & RMSpro             & 10              \\ 
\textbf{\#4}               & Thursday         & Web Attack: Brute Force, XSS, and SQL Injection         & 0.00013936442920558617                                                 & Adam               & 10              \\ 
\textbf{\#5}               & Friday           & Firewall Rules: On and Off                                                                                                                                                & 0.000587441102433820                                                    & RMSprop            & 10              \\ 
\textbf{\#6}               & Friday           & Botnet Ares                                                                                                                                                               & 0.0006052967400865347                                                  & SGD                & 10              \\ 
\textbf{\#7}               & Friday           & DDoS LOIT                                                                                                                                                                 & 0.00012091571705782663                                                 & Adam               & 10              \\ \hline 
\end{tabular}%
\end{table*}

Through the process of optimizing the hyperparameters, we were able to ensure that each \texttt{ML-Agent} could train effectively on its respective local dataset. Subsequently, we evaluated the training capacity and performance of each \texttt{ML-Agent} using their local dataset. The \texttt{ML-Agent} column indicates the training agent employed during the experiment. In contrast, the ``Dataset'' column provides details on the data assigned to the \texttt{ML-Agent}, and the ``Attacks'' column specifies the types of attacks the \texttt{ML-Agent} was trained to classify/identify.

\subsubsection{Training Behavior}\label{subsubsec:training_behaviour}

\begin{figure}[tbp]
  \centering
  \includegraphics[width=0.85\columnwidth]{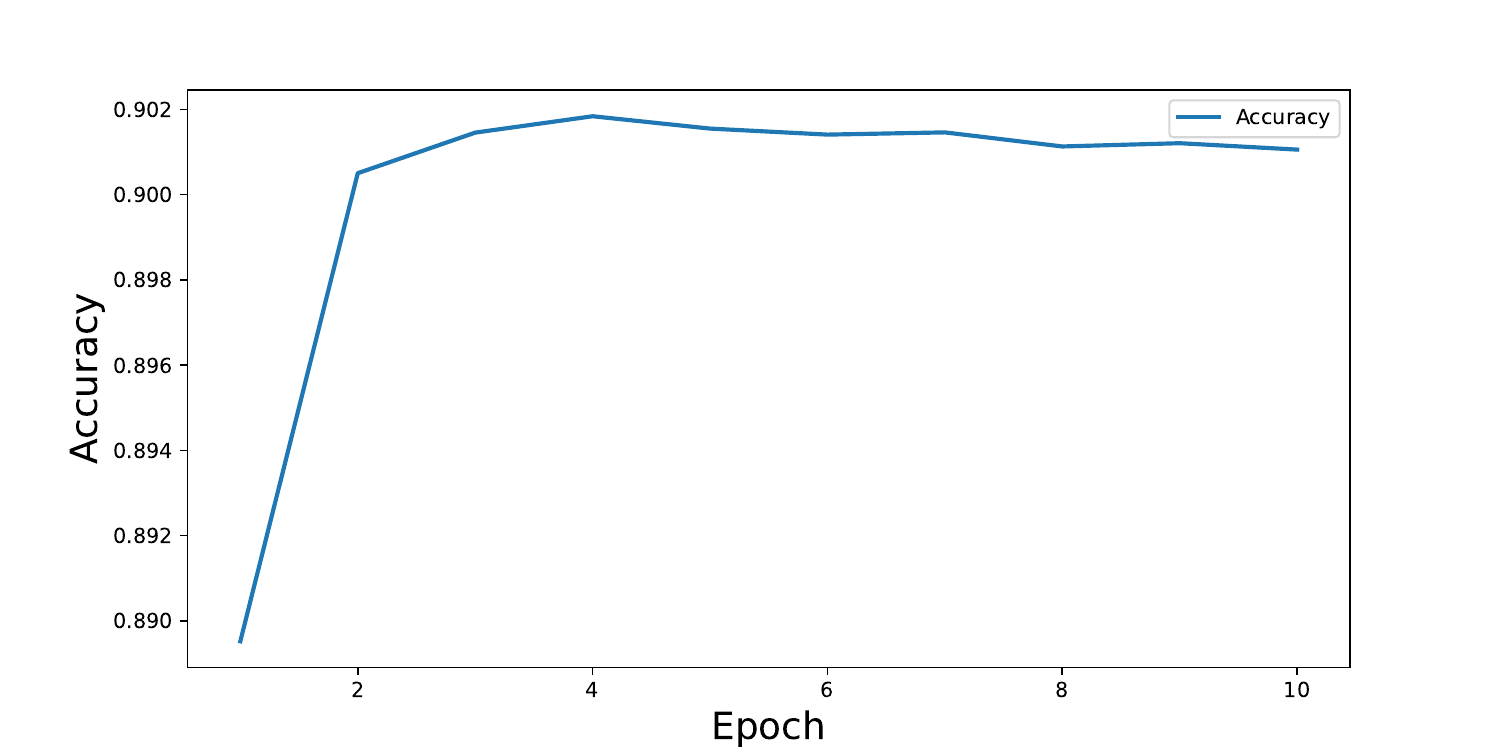}
  \caption{Accuracy behavior for a joint dataset using a centralized training approach.}
  \label{fig:accuracies_for_centralized_training}
\end{figure}

\begin{figure}[tbp]
  \centering
  \includegraphics[width=0.85\columnwidth]{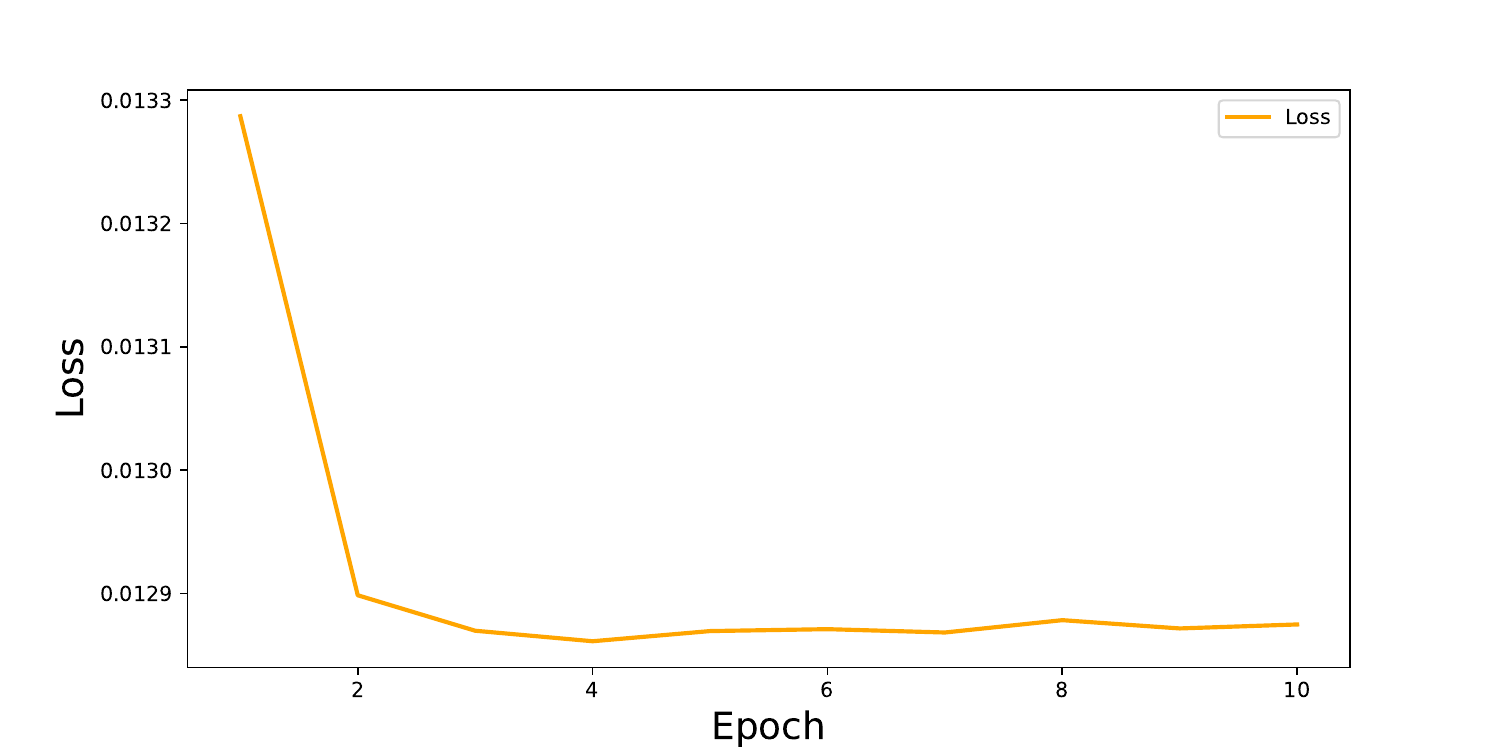}
  \caption{Loss behavior for a joint dataset using a centralized training approach.}
  \label{fig:loss_for_centralized_training}
\end{figure}

We evaluated the training performance of a neural network using binary classification of malignant or benign traffic. Our native \ac{AI} and security architecture are flexible enough to support different types of approaches for handling threats defense and training \ac{AI} models to handle security. Initially, we validated two behaviors of SFI2 \ac{AI} Management, triggering centralized training or distributed training across \texttt{ML-Agents} coupled as microservices in the architecture. For centralized, we grouped the seven datasets in this experimental scenario, leading to a centralized training approach. Thus, after ten (10) different runs, we obtained an average test accuracy of 90.01\%. We ensured that learning was consistent by presenting the loss function and training accuracy graphs in  Fig.~\ref{fig:accuracies_for_centralized_training}, and Fig.~\ref{fig:loss_for_centralized_training}. 

Although visually, the Accuracy and Loss graphs in Fig.~\ref{fig:accuracies_for_centralized_training} and Fig.~\ref{fig:loss_for_centralized_training} appear to fluctuate slightly; in our experiments, the model converged at epoch 10 when there were no more significant gains in accuracy, and we activated the early stopping of the learning process. It should also be noted that the aggregation of the dataset culminated in a dataset with higher CPU and memory consumption, and training took an average of 1072 seconds.

\begin{figure*}[!h]
  \centering
  \includegraphics[width=1\textwidth]{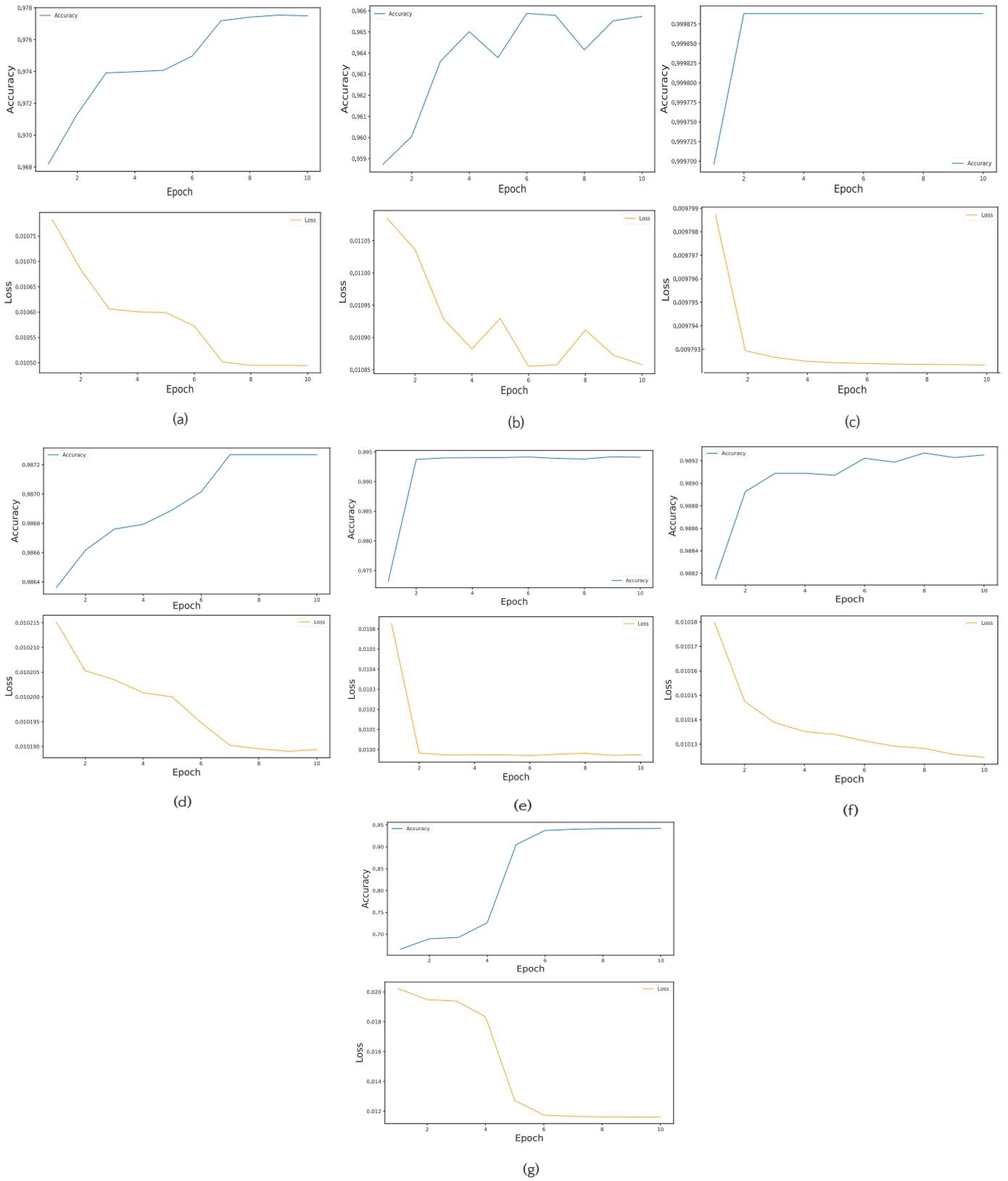}
  \caption{Accuracies for each \texttt{ML-Agent}. The graphs \textit{(a)} to \textit{(g)} refer to the accuracy and loss behaviors of each \texttt{ML-Agent} with its local dataset according to Table~\ref{tab:hyperparameters_values}.}
  \label{fig:accuracies_for_each_client}
\end{figure*}

Subsequently, we analyzed in Fig.~\ref{fig:accuracies_for_each_client} the behavior of the accuracy and loss curves for the federated learning scenario involving two training rounds. This distributed scenario aligns with our contribution because it enables each \texttt{Security Agent} to deal with potential security threats as a specific entity in the network slicing architecture. Biases in weight aggregation averages can compromise centralized AI models. 

As  Fig.~\ref{fig:accuracies_for_each_client} shows, each model exhibited different behaviors during the training process. However, it is worth noting that the \texttt{ML-Agents} showed appropriate behavior in the curves, indicating that they could converge in learning. Consequently, the server model achieved an average accuracy of 90.8\% using two training rounds. The variation in accuracy and loss levels between the different \texttt{\ac{ML}-Agents}, as shown in Fig.~\ref{fig:accuracies_for_each_client}, is an intrinsic characteristic of federated learning in  a \textit{non}-\ac{IID} scenario, where each \texttt{\ac{ML}-Agent} can have a unique learning behavior over the time.

In addition, as shown in Fig.~\ref{fig:accuracies_for_each_client} graphs represent how challenging it was for \texttt{\ac{ML}-Agents} to deal with \textit{non}-\ac{IID} data while creating a global \ac{AI} model to empower network slicing architectures, dealing with different types of \ac{DDoS} and intrusion attacks. Although challenging, the models of each \texttt{\ac{ML}-Agent} converged in learning because the stabilization of the loss and accuracy after the end of the epochs was noted. With this, we have that each \texttt{\ac{ML}-Agent} dealt with different types of \ac{DDoS} attacks and, at the same time, contributed to adjusting a generic and robust \ac{AI} model for the architecture of networking slicing.

\begin{figure}[htpb]
  \centering
  \includegraphics[width=.6\columnwidth]{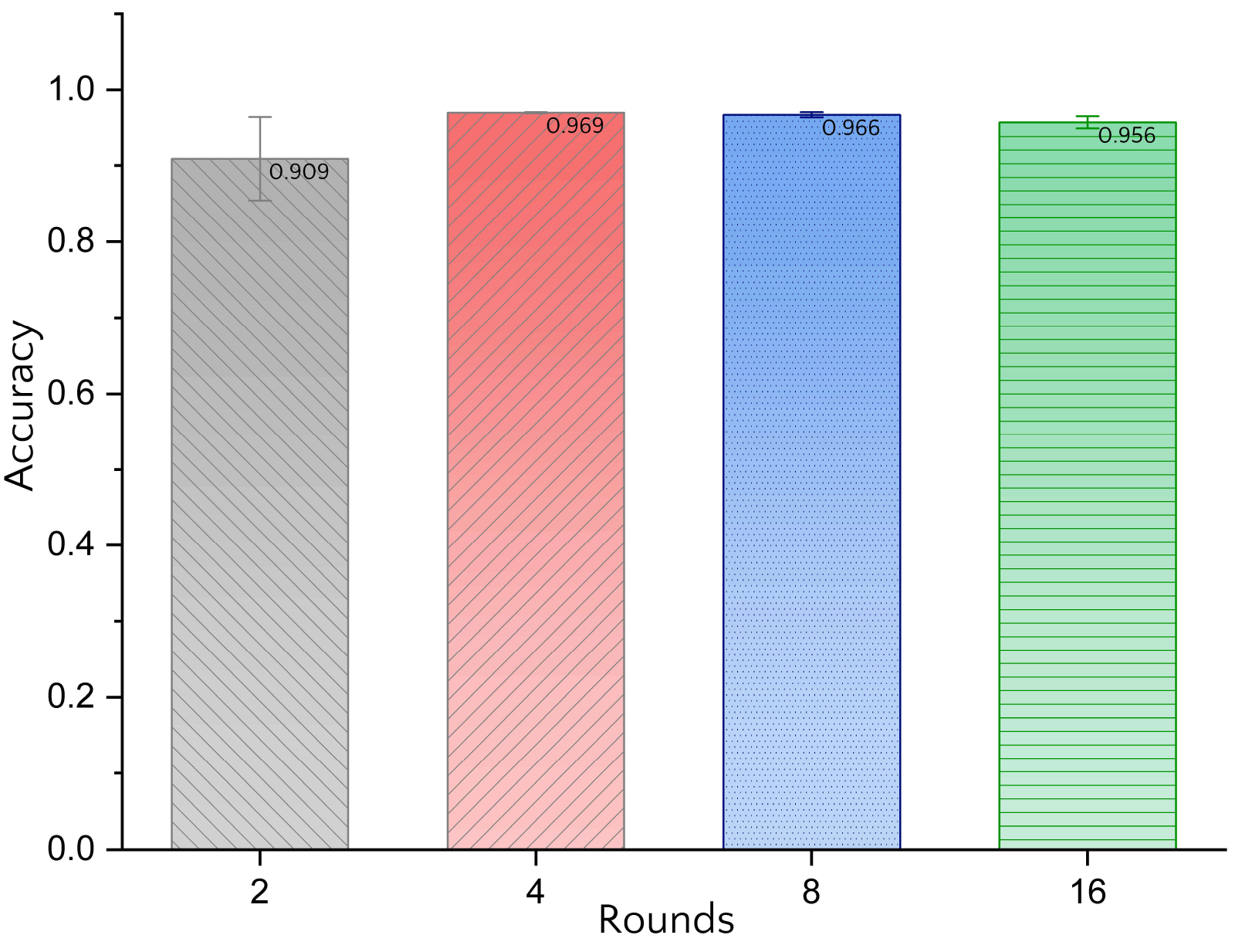}
  \caption{Accuracy for different Federated Rounds.}
  \label{fig:accuracy_over_each_dataset}
\end{figure}

\begin{table*}[htbp]
\centering
\caption{Comparison of training time considering two different approaches.}
\label{tab:training_time_comparison}
\scriptsize
\begin{tabular}{ccccccccccc}
\hline
\textbf{Approach}            & \textbf{Round}       & \textbf{Count}       & \textbf{\begin{tabular}[c]{@{}c@{}}Mean \\ Training Time\end{tabular}} & \textbf{\begin{tabular}[c]{@{}c@{}}Standard Error \\ Mean\end{tabular}} & \textbf{\begin{tabular}[c]{@{}c@{}}Standard \\ Deviation\end{tabular}} & \textbf{Minimum} & \textbf{Q1} & \textbf{Median} & \textbf{Q3} & \textbf{Maximum} \\ \hline
\multirow{4}{*}{Federated}   & Two                  & 280                  & 362.37                                                                 & 6.43             & 107.57         & 236.66           & 257.7       & 344.06          & 452.29      & 619.3            \\
                             & Four                 & 420                  & 358.55                                                                 & 5.22             & 107            & 235.57           & 255.5       & 338.67          & 446.07      & 622.77           \\
                             & Eight                & 980                  & 358.9                                                                  & 3.4              & 106.34         & 235.57           & 256.8       & 339.78          & 446.51      & 622.77           \\
                             & Sixteen              & 1832                 & 378.64                                                                 & 2.65             & 113.24         & 235.57           & 277.29      & 344.95          & 453.35      & 628.73           \\ \hline
\multirow{2}{*}{Centralized} & \multicolumn{1}{l}{} & \multicolumn{1}{l}{} & \textbf{\begin{tabular}[c]{@{}c@{}}Mean\\ Training Time\end{tabular}}  & \textbf{SE Mean} & \textbf{StDev} & \textbf{Minimum} & \textbf{Q1} & \textbf{Median} & \textbf{Q3} & \textbf{Maximum} \\
                             & \multicolumn{1}{l}{} & 10                   & 1072.3                                                                 & 0.249            & 0.789          & 1070.5           & 1072.3      & 1072.5          & 1072.5      & 1073.5           \\ \hline
\end{tabular}%
\end{table*}

After conducting a thorough analysis of the capabilities of the \texttt{ML-Agents} to train federatively with \textit{non}-IID datasets and reporting the weights to the central model in the SFI2 Architecture, we examined the effect of federated learning rounds on accuracy. Fig.~\ref{fig:accuracy_over_each_dataset} shows the variability of the aggregate accuracy of the model concerning different interaction rounds. Our findings indicate that, with an error of less than $5\%$, increasing the number of training rounds had no significant influence on the accuracy of the central model. It can, therefore, be deduced that the \textit{non}-IID datasets may not benefit from long federated training rounds.

We present a comparison of the training times for the centralized and federated approaches in Table~\ref{tab:training_time_comparison}. Subsequently, we analyzed the models resulting from federated clients with the aggregated server model by utilizing cosine divergence to assess the significance of any differences between the server and client models (vector of weights) across various training rounds. A substantial difference suggests that a particular customer may be overlooked because of its minimal impact on the convergence of the model. The average cosine divergence between client models and the server is presented in Table~\ref{tab:descriptive_stats}.

\begin{table}[htbp]
\centering
\caption{Descriptive Statistics of cosine divergence for each training round.}
\label{tab:descriptive_stats}
\scriptsize
\begin{tabular}{ccccc}
\cline{2-5} 
\multicolumn{1}{l}{} & \textbf{N Analysis} & \textbf{Mean} & \textbf{Standard Deviation} & \textbf{SE of Mean} \\ \hline \hline
2                    & 7                   & 0.00446       & 0.02164                     & 0.00818             \\
4                    & 7                   & -0.00045      & 0.02043                     & 0.00772             \\
8                    & 7                   & 0.01440        & 0.01721                     & 0.00650             \\
16                   & 7                   & 0.01746       & 0.02376                     & 0.00898             \\ \hline \hline
\end{tabular}%
\end{table}

\subsubsection{Analysis of Models}\label{subsubsec:models_analyses}

Regarding the training paradigm analyses, we compared the models resulting from federated clients with the aggregated server model by utilizing cosine divergence to assess the significance of any differences between the client and server models across various training rounds. Upon obtaining the samples and the average cosine divergence between the client models and server, we analyzed the variance (ANOVA) to evaluate the statistical equivalence of these samples. We employed an ANOVA with four levels, each representing a sample of the cosine difference for the different training rounds. We formulate the following hypotheses for our analysis:

\begin{itemize}
    \item \textit{Null Hypothesis}: The means of all levels are equal.
    \item \textit{Alternative Hypothesis}: The means of one or more levels are different.
\end{itemize}

The ANOVA test results presented in Table~\ref{tab:anova_results}, especially the \textit{p}-value, indicate no statistically significant difference between the means of the four variables; namely, increasing the number of training rounds in federated learning does not affect the accuracy achieved. Specifically, the value of \textit{p} is 0.35479, indicating that we should accept the null hypothesis that the means are equal at a significance level of $5\%$.

According to Table~\ref{tab:r-score}, the R-squared value is 0.12427, which indicates that the model accounts for only $12.43\%$ of the variation in the data. Hence, it can be inferred that other factors, such as the size of the dataset, the type of algorithm employed, and the quality of communication between nodes, significantly influence the model's accuracy beyond the number of training rounds.

\begin{table}[htbp]
\centering
\caption{Analysis of Variance Test.}
\label{tab:anova_results}
\scriptsize
\begin{tabular}{cccccc}
\cline{2-6}
\multicolumn{1}{l}{} & \textbf{DF} & \textbf{Sum of Squares} & \textbf{Mean Square} & \textbf{F Value} & \textbf{Prob$>$F} \\ \hline \hline
\textbf{Model}       & 3           & 0.00149                 & 0.00050              & 1.13519          & 0.35479         \\
\textbf{Error}       & 24          & 0.01048                 & 0.00044              &                  &                 \\
\textbf{Total}       & 27          & 0.01196                 &                      &                  &                 \\ \hline \hline
\end{tabular}%
\end{table}

\begin{table}[ht]

\centering
\caption{Fit Statistics.}
\scriptsize
\label{tab:r-score}
\begin{tabular}{cccc}
\hline
\textbf{R-Square} & \textbf{Coeff Var} & \textbf{Root MSE} & \textbf{Data Mean} \\ \hline
0.12427           & 2.33603            & 0.02089           & 0.00894            \\ \hline
\end{tabular}%
\end{table}

\begin{figure}[H]
  \centering
  \includegraphics[width=0.8\columnwidth]{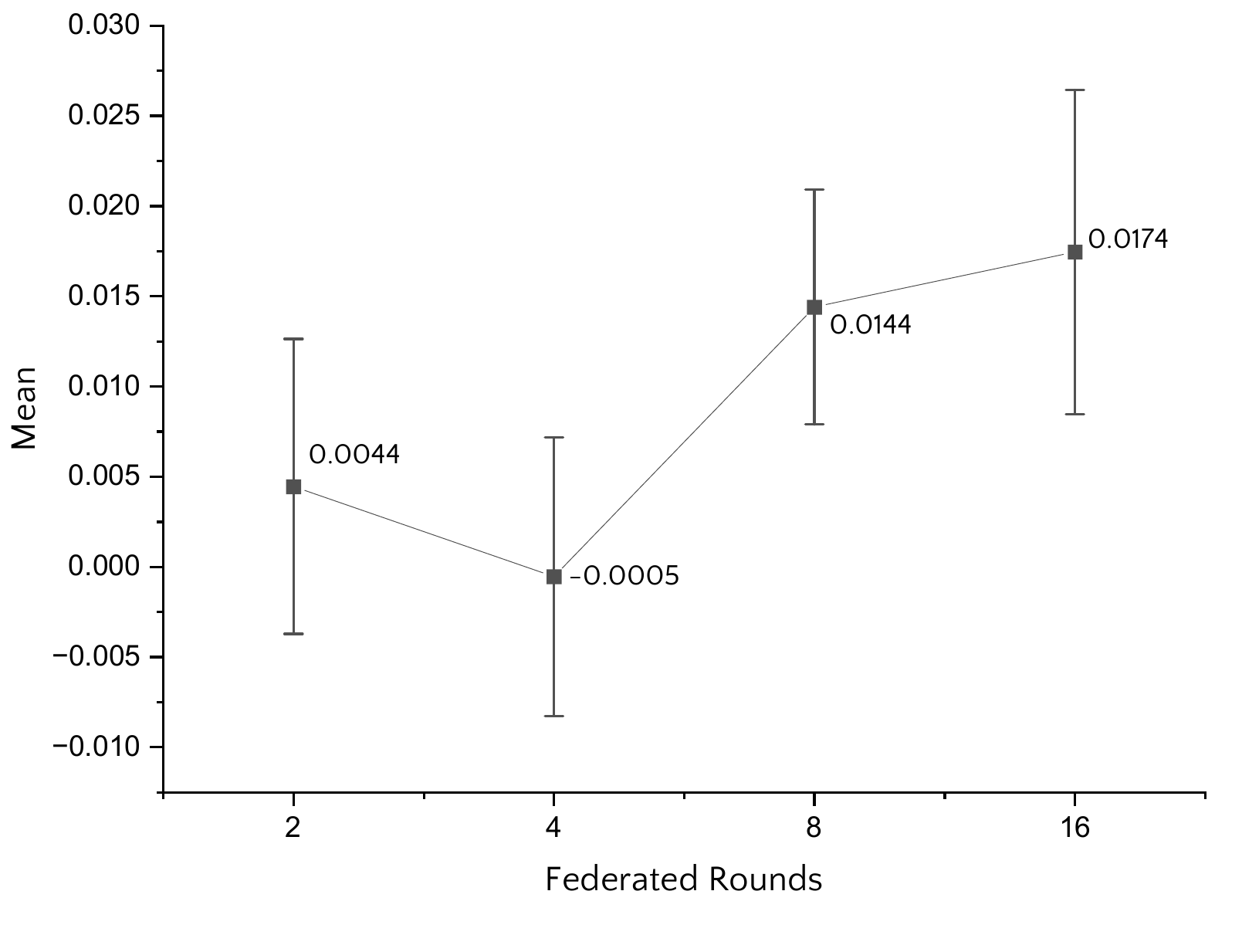}
  \caption{Cosine Variation through different training Rounds.}
  \label{fig:cosine_variation}
\end{figure}

Fig.~\ref{fig:cosine_variation} shows the statistical equivalence between the cosine differences and Standard Error (SE) of the mean according to the results of the ANOVA test. This implies that the model can learn and improve, albeit without statistically significant improvements. Additionally, the model seems to converge towards a solution as it progresses towards distributed learning.

The Table~\ref{tab:ns_slicing_sota_comparison} presents a comparative analysis of various state-of-the-art approaches in addressing different security threats using distinct datasets and employed methods. Each approach is evaluated based on its capability to support multiple network slicing (NS) architectures, provide proactive monitoring, implement low-overhead monitoring, and its overall defense efficiency. The results indicate that the accuracy or precision of these methods varies. Our proposed approach, utilizing the CIC-IDS2017 dataset with an \ac{LSTM} model, demonstrates competitive acuracy at 96.60\%, while also meeting the criteria for multiple NS architecture support, proactive monitoring, and low-overhead monitoring.

\begin{table}[H]
\centering
\caption{Comparison of \ac{FL}-based Security Defense Mechanisms for \ac{NS} Architectures.}
\label{tab:ns_slicing_sota_comparison}
\Huge
\resizebox{\columnwidth}{!}{%
\begin{tabular}{|c|c|c|c|c|c|c|c|}
\hline
\textbf{Approach}                                                                           & \textbf{Dataset}                                & \textbf{\begin{tabular}[c]{@{}c@{}}Security \\ Threat\end{tabular}}            & \textbf{\begin{tabular}[c]{@{}c@{}}Employed \\ Method\end{tabular}} & \textbf{\begin{tabular}[c]{@{}c@{}}Multiple \\ NS Architecture\\ Support\end{tabular}} & \textbf{\begin{tabular}[c]{@{}c@{}}Pro-active \\ Monitoring\end{tabular}} & \textbf{\begin{tabular}[c]{@{}c@{}}Low-overhead \\ Monitoring\end{tabular}} & \textbf{\begin{tabular}[c]{@{}c@{}}Defense \\ Efficiency\end{tabular}} \\ \hline
Niboucha et. al~\cite{Niboucha2023}                                                         & Own                                             & \ac{DDoS}                                                                        & Gradient Boosting                                                   & \faCircleO                                                                             & \faCircleO                                                                & \faCircleO                                                                  & \begin{tabular}[c]{@{}c@{}}Accuracy: \\ $96.76\%$\end{tabular}         \\ \hline
Wijethilaka et. al~\cite{Wijethilaka2022}                                                   & NSL-KDD~\cite{nsl-kdd}                          & \begin{tabular}[c]{@{}c@{}}\ac{DoS}, and \\ \ac{U2R}\end{tabular}                & \ac{DL}                                                               & \faCircleO                                                                             & \faCircle                                                                 & \faCircleO                                                                  & \begin{tabular}[c]{@{}c@{}}Accuracy: \\ $99.99\%$\end{tabular}         \\ \hline
\begin{tabular}[c]{@{}c@{}}Sedjelmaci and \\ Boualouache~\cite{Sedjelmaci2024}\end{tabular} & CSE-CIC-IDS-2018~\cite{Sharafaldin2018TowardGA} & \ac{DDoS} and Botnet                                                                & Mean-field Game                                                     & \faCircleO                                                                             & \faCircleO                                                                & \faCircleO                                                                  & \begin{tabular}[c]{@{}c@{}}Accuracy: \\ $97.00\%$\end{tabular}         \\ \hline
Wijethilaka et. al~\cite{Wijethilaka2023}                                                   & NSL-KDD~\cite{nsl-kdd}                          & \begin{tabular}[c]{@{}c@{}}\ac{DDoS}, \\ \ac{MITM}, \\ and Botnet\end{tabular} & \ac{DL}                                                               & \faCircleO                                                                             & \faCircle                                                                 & \faCircleO                                                                  & \begin{tabular}[c]{@{}c@{}}Accuracy: \\ $98.00\%$\end{tabular}         \\ \hline
Rumesh et. al~\cite{Rumesh2024}                                                             & Own                                             & \ac{DDoS}                                                                        & \ac{LSTM}                                                             & \faCircleO                                                                             & \faCircle                                                                 & \faCircleO                                                                  & \begin{tabular}[c]{@{}c@{}}Accuracy: \\ $99.87\%$\end{tabular}         \\ \hline
Mirzaee et. al~\cite{Mirzaee2021}                                                           & NSL-KDD~\cite{nsl-kdd}                          & \ac{DoS} and \ac{U2R}                                                               & \ac{DL}                                                               & \faCircleO                                                                             & \faCircle                                                                 & \faCircleO                                                                  & \begin{tabular}[c]{@{}c@{}}Accuracy: \\ $99.50\%$\end{tabular}         \\ \hline
Thantharate~\cite{Thantharate2022}                                                          & Own                                             & \ac{DDoS}                                                                        & Sequential Model                                                    & \faCircleO                                                                             & \faCircleO                                                                & \faCircleO                                                                  & \begin{tabular}[c]{@{}c@{}}Precision: \\ $94.00\%$\end{tabular}        \\ \hline
\textbf{Our Approach}                                                                       & CIC-IDS2017~\cite{Sharafaldin2018}              & \ac{DDoS}                                                                        & \ac{LSTM}                                                             & \faCircle                                                                              & \faCircle                                                                 & \faCircle                                                                   & \begin{tabular}[c]{@{}c@{}}Accuracy: \\ $96.60\%$\end{tabular}        \\ \hline
\end{tabular}%
}
\end{table}

\section{Concluding Remarks}\label{sec:concluding_remark}

In this paper, we presented a microservice-based approach to enhance the intelligence and security of network-slicing architectures. Our research reveals that several network slicing architectures lack the necessary intelligence and security features to effectively operate and protect network slices. 
To address this shortcoming, we propose using \texttt{ML-Agents} and \texttt{Security Agents}, which collaborate to provide intelligent and secure management and orchestration for network slice core entities.

Our research demonstrated that federated learning, when associated with microservice architectures, can enhance network-slicing architectures, improve their resilience to various security attacks, and build robust \ac{AI} models for attack prediction for architecture blocks and intra-slices. We conclude that federated learning can make an architecture more resilient to threats, mainly by providing on-demand adjustments and adapting to changing data. In addition, we evaluated the behavior of training rounds in federated learning. We determined that the number of training rounds was insignificant for constructing these \ac{AI} models.

The results of this study offer new insights into the evolution of architectures and frameworks for network slicing, allowing their extension to the design of architectures adapted to security that support the requirements of new applications and business verticals.

For future work, we plan to focus on enhancing our solution by integrating new machine learning models to bolster its capacity to respond to security threats. Additionally, we aim to explore the application of reinforcement learning to ensure the robustness of network slicing architectures, even in novel attack scenarios. We also intend to investigate the impact of different security methods on service-level agreements and operator revenue.

This study reports recent advancements and highlights significant research opportunities in intelligent and security-aware resource sharing for future network architectures.

\section*{Acknowledgements}

We acknowledge the financial support of the Brazilian National Council for Scientific and Technological Development (CNPq), grant \#421944/2021-8 and the FAPESP MCTIC/CGI Research project 2018/23097-3 - SFI2 - Slicing Future Internet Infrastructures. The authors also thanks CNPq, CAPES, FAPES and Instituto ANIMA.





\bibliographystyle{elsarticle-num} 
\bibliography{references}



\end{document}